\documentclass[twocolumn, nofootinbib, aps, prd, 10pt]{revtex4-1}
\usepackage{graphicx}
\usepackage{float}
\usepackage{url}
\usepackage{color}
\usepackage{rotating}
\usepackage{amssymb,amsmath}
\usepackage{mathtools} 

\newcommand{\Ylm}{Y_{\ell m}}
\newcommand{\tlm}{t_{\ell m}}
\newcommand{\clm}{c_{\ell m}}
\newcommand{\alm}{a_{\ell m}}
\newcommand{\glm}{g_{\ell m}}

\newcommand{\Cl}{C_\ell}

\newcommand{\nhat}{{\bf \hat n}}

\newcommand\myeq{\stackrel{\mathclap{\normalfont\mbox{\scriptsize single LSS}}}{\hspace{0.45cm}\longrightarrow\hspace{0.45cm}}}
\newcommand\myeqtwo{\stackrel{\mathclap{\normalfont\mbox{\scriptsize large A}}}{\hspace{0.45cm}\longrightarrow\hspace{0.45cm}}}

\everymath{\displaystyle}
  
\newcommand{\be}{\begin{equation}}
\newcommand{\ee}{\end{equation}}

\begin{document}

\title{
Reconstructing the integrated Sachs-Wolfe map with galaxy surveys}

\author{Jessica Muir and Dragan Huterer}
\affiliation{Department of Physics, University of Michigan, 
450 Church St, Ann Arbor, MI 48109-1040}

\date{\today}

\begin{abstract}
The integrated Sachs-Wolfe (ISW) effect is a large-angle modulation of the
cosmic microwave background (CMB), generated when CMB photons traverse
evolving potential wells associated with large scale structure (LSS). Recent
efforts have been made to reconstruct maps of the ISW signal using information
from surveys of galaxies and other LSS tracers, but investigation into how
survey systematics affect their reliability has so far been limited.  Using
simulated ISW and LSS maps, we study the impact of galaxy survey properties
and systematic errors on the accuracy of reconstructed ISW signal. We find
that systematics that affect the observed distribution of galaxies along the
line of sight, such as photo-$z$ and bias-evolution related errors, have
a relatively minor impact on reconstruction quality.
  In contrast, however, we  find 
that direction-dependent calibration errors can be very harmful.
Specifically, we find that in order to avoid significant degradation of our
reconstruction quality statistics, direction-dependent number density
fluctuations due to systematics must be controlled so that their variance is
smaller than $10^{-6}$ (which corresponds to a 0.1\% calibration).
Additionally, we explore the implications of our results for attempts to use
reconstructed ISW maps to shed light on the origin of large-angle CMB
alignments. We find that there is only a weak correlation between the true and
reconstructed angular momentum dispersion, which quantifies alignment, even
for reconstructed ISW maps which are fairly accurate overall.

\end{abstract}
\maketitle
\section{Introduction}\label{sec:intro}

As cosmic microwave background (CMB) photons travel from the last scattering
surface to our detectors, they can experience a frequency shift beyond that
which is guaranteed by the expansion of the universe.
This additional effect is a result of the fact that gravitational potential
fluctuations associated with large-scale structure (LSS) decay with time when
the universe is not fully matter dominated.  Consequently, the CMB photons are
subject to a direction-dependent temperature modulation which is proportional
to twice the rate of change in the potential integrated along the line of
sight. This modulation is known as the integrated Sachs-Wolfe (ISW)
effect~\cite{Hu:1993xh}. Its magnitude in direction $\nhat$ on the sky was
worked out in the classic Sachs-Wolfe paper~\citep{Sachs:1967er} to be
\begin{equation}\label{ISWorigexpr}
\left.\frac{\Delta T}{\bar{T}}\right|_{ISW}(\nhat) = \frac{2}{c^2}\int_{t_*}^{t_0}dt\,\frac{\partial \Phi(\mathbf{r},t)}{\partial t},
\end{equation}
where $t_0$ is the present time. $t_{\star}$ is that of recombination, $c$ is the speed of light, $\mathbf{r}$ is the position in comoving coordinates, and
$\Phi$ is the gravitational potential.

The ISW effect introduces a weak additional signal at very large scales (low
multipoles) in the CMB angular power spectrum. It carries important
information about dark energy \cite{Bean:2003fb,Weller:2003hw}, particularly
its clustering properties that are often parametrized by the dark energy speed
of sound. It also potentially offers useful information about the the nature
of dark energy, as modified gravity theories have unique ISW signatures
\cite{Song:2006jk}. However, the fact that the largest CMB multipoles are
subject to cosmic variance severely limits how much information can be gleaned
from the ISW given the CMB temperature measurements alone.

We are able to observe the ISW effect because the dependence of the ISW signal
on the time derivative of the potential results in a large-angle cross-correlation
between LSS tracers and CMB temperature.  This was first pointed
out by Crittenden \& Turok \cite{Crittenden:1995ak}, who further suggested
cross-correlation between CMB temperature anisotropy $(\delta T/T)_{\rm
  ISW}({\bf \hat{n}}) $ and galaxy positions, $(\delta N/N) ({\bf \hat{n}'})$,
as a statistic through which to detect the ISW effect. This
  cross-correlation signal was detected shortly
thereafter \citep{Boughn:2003yz} and was later confirmed by many teams who
found cumulative evidence of about $4\sigma$ using a number of
different LSS tracers \cite{Fosalba:2003ge, Nolta:2003uy, Corasaniti:2005pq,
  Padmanabhan:2004fy, Vielva:2004zg, McEwen:2006my,Giannantonio:2006du,
  Cabre:2007rv, Rassat:2006kq,
  Giannantonio:2008zi,Ho:2008bz,Xia:2009dr,Giannantonio:2012aa,Ade:2013dsi,Ade:2015dva}.
Comprehensive surveys of recent results can be found
in Refs.~\cite{Dupe:2010zs,Giannantonio:2012aa,Ade:2015dva}. While the detection of
the ISW effect itself provides independent evidence for dark energy at high
statistical significance, prospects for using it to constrain the cosmological
parameters are somewhat limited \citep{Hu_Scranton}.

The ISW {\it map}, $(\delta T/T)_{\rm ISW}({\bf \hat{n}})$, is also of
interest in its own right. By assuming a cosmological model, one can construct
an estimator using theoretical cross-correlations in combination with LSS
data.  Because the ISW signal represents a late-universe contribution to the
CMB anisotropy, measuring and subtracting it from observed temperature
fluctuations would allow us to isolate the (dominant) early-universe
contributions to the CMB.  If this procedure could be done reliably, it would
have immediate implications for our understanding of the cosmological model.

For example, the ISW signal has been identified as a potential contributor to
large-angle CMB features which have been reported to be in tension with the
predictions of $\Lambda$CDM~\cite{Schwarz:2015cma}.  A reconstructed ISW map
would clarify whether some component of the CMB anomalies (discussed further
below in Sec.~\ref{sec:anomalies}) become stronger or weaker when evaluated on
the early-universe-only contribution to the CMB. A few
studies~\cite{Francis:2009pt,Rassat:2013caa} have already explored this.  To
study the impact of ISW contributions on CMB anomalies,
Ref.~\cite{Rassat:2013caa} uses WMAP data with 2MASS and NVSS, while
Ref.~\cite{Francis:2009pt} uses 2MASS alone.

The late-time ISW also provides a contaminant to the measurement of 
primordial non-Gaussianity from CMB maps. Because both the ISW effect and
gravitational lensing trace LSS, they couple large- and small-scale modes of
the CMB, resulting in a nonprimordial contribution to the bispectrum.
Recent analyses~\cite{Ade:2015ava} have corrected for this by
including a theoretical template for the ISW-lensing bispectrum in primordial
$f_{NL}$ analyses. Reconstructing and subtracting the ISW contribution from
the CMB temperature maps could provide an alternative method for removing
ISW-lensing bias when studying primordial non-Gaussianity~\cite{Kim:2013nea}.

More generally, understanding how reliably the ISW map can be reconstructed
from large-scale structure information impacts our understanding of 
how the late universe affects our view of the primordial CMB sky.

Before reconstruction can be done reliably, however, we must understand how
systematics associated with the input data impact the ISW estimator's
accuracy. Previous works have explored this to some extent, looking at how
reconstruction quality is affected by the inclusion of different input
data sets~\cite{Manzotti:2014kta,Ade:2015dva,Bonavera:2016hbm},
masks~\cite{Ade:2015dva,Bonavera:2016hbm} and, to a limited degree, the
influence of uncertainties in cosmological and bias
models~\cite{Bonavera:2016hbm}. Additionally, Ref.~\cite{Afshordi:2004kz} studied how systematics like redshift uncertainties and photometric calibration change the signal to noise of the ISW effect's {\it detection}. That being said, there remain a number of
systematics inherent to galaxy survey data which have not yet been
subject to detailed analysis in the context of ISW map reconstruction. We aim to address this.

In this paper, we use simulated ISW and LSS maps to
identify which survey properties are important for ISW reconstruction and to
quantify their effects on the reconstructed maps. We begin by studying how
survey depth, redshift binning strategy, and the minimum measured multipole
$\ell_{\rm min}$ influence reconstruction quality in the absence of
systematics.  Using these results as a baseline, we then explore two broad
classes of systematics: ways one can mismodel the redshift distribution of
LSS sources, 
and direction-dependent photometric calibration errors that can result from,
for example, contamination by stars. We also briefly discuss the implications
of our results for analysis of whether the ISW signal contributes to the
observed alignments between large-angle multipoles of the CMB temperature map.

The paper is organized as follows. In Sec.~\ref{sec:methods} we discuss our 
general procedure for the ISW map reconstruction and assessment of the
accuracy in this procedure. In Sec.~\ref{sec:surveyprops}, we describe the
properties of the surveys that we will consider, while in
Sec.~\ref{sec:surveysys}, we discuss the effect of various systematic errors on
the ISW map reconstruction. We conclude in Sec.~\ref{sec:concl}.

\section{Methods}
\label{sec:methods}
We perform a number of studies examining how  survey properties and
systematics affect the accuracy of reconstructed ISW maps. These studies all
follow this general pipeline:
\begin{itemize}
  \item Select a fiducial cosmological model and 
    specifications of the LSS survey.
\item Compute the ``true'' angular cross-power $\Cl^{XY}$ for ISW and LSS maps,
  assuming the fiducial cosmology and survey specifications.
\item Use the true $\Cl^{XY}$ to generate correlated Gaussian
  realizations of the true ISW signal and corresponding LSS maps.
\item If applicable, postprocess the galaxy maps to model direction-dependent
  systematic effects. 
\item Construct an estimator for the ISW signal using the simulated galaxy
  maps and a set of ``model'' $\Cl^{XY}$ which may or
  may not match those used to generate the simulations.
\item Compare the reconstructed ISW signal to the true ISW map and
evaluate the accuracy of the reconstruction.
\end{itemize}

This section will introduce some of the theoretical tools needed for this analysis.

\begin{figure}
\includegraphics[width=\linewidth]{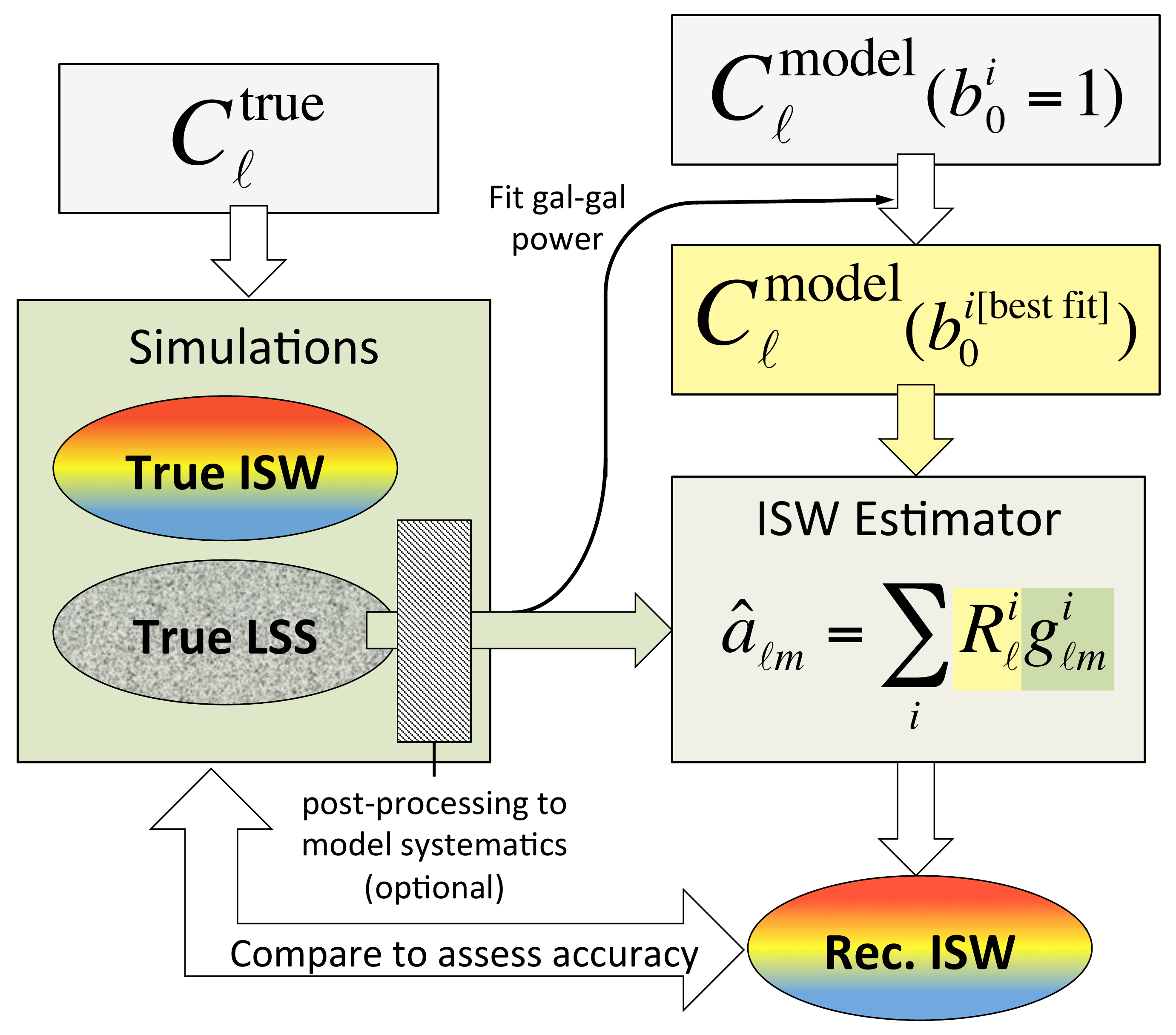}
  \caption{Flowchart of reconstruction pipeline.}
  \label{fig:pipelineflowchart}
\end{figure}

\subsection{Theoretical cross-correlations}\label{sec:clxy}

The angular cross-power between ISW and galaxy maps serves as input for both
the simulation and reconstruction processes used in the following sections. Given
maps $X$ and $Y$, the expression for the angular cross-power between them is
\be
\Cl^{XY} = \frac{2}{\pi}\int dk\,k^2\,P(k)\,I_{\ell}^X(k)\,I_{\ell}^Y(k)
\ee
where $P(k)$ is the matter power spectrum at $z=0$, and the transfer function
$I_{\ell}^X(k)$ is written 
\be\label{eq:Ilk_transferfunc}
I_{\ell}^X(k) \equiv \int_0^\infty\,dz\,D(z)\,W^X(z,k)\,j_{\ell}(kr).
\ee
Here, $r\equiv r(z)$ represents comoving radius; $j_{\ell}(x)$ is a spherical
Bessel function; and $D(z)$, which is normalized to one at $z=0$,  describes
the linear growth of matter fluctuations. The function $W^X(k,z)$ is a
tracer-specific window function that encapsulates the relationship between the
tracer $X$ and underlying dark matter fluctuations $\delta$.
The tracers relevant to our studies are the ISW signal and  galaxy number density. 

The ISW window function is
\begin{equation}
  W^{\rm ISW}(z,k) = \left[\Theta(z_{\rm max}-z)\right]\,
  \left[\frac{3H_0^2\Omega_m}{c^2k^2}\right]\left(1-f(z)\right),
\end{equation}
where $\Theta$ is the Heaviside step function.  In this expression, the term in
square brackets comes from when the Poisson equation is used to relate
potential fluctuations to dark matter density, $\Omega_m$ is the
  matter density in units of the critical density, and $H_0$ is the
present-day Hubble parameter. The appearance of the
growth rate $f(z)\equiv d\ln D/d\ln a$ comes from the time derivative in
Eq.~(\ref{ISWorigexpr}). To compute the full ISW contribution, one would
integrate to the redshift of recombination, $z_{\rm max}=z_{\star}$.  In this
work, though, we are interested only in the late ISW effect, so we can set
$z_{\rm max}=15$ without a loss in accuracy.

Each survey (and each redshift bin within a given survey) will have its own
window function.  For a map of galaxy number
  density fluctuations, it is  
\be\label{eq:windowgal}
  W^{\rm gal}(z,k) = b(z)\,\frac{dn}{dz}.
\ee
In this expression, $b(z)$ represents linear bias, which we assume is
scale independent. The function $dn/dz$ describes the redshift distribution
the observed sources, encapsulating information about how their physical
density varies with redshift as well as survey volume and selection effects.
It is normalized so it integrates to one.
Galaxy shot noise is included by adding a contribution to its autopower spectrum,
\be\label{eq:shotnoise}
\Cl^{\rm gal-gal}\rightarrow \Cl^{\rm gal-gal} + \bar{n}^{-1}
\ee
where $\bar{n}$ is the average number density of sources per steradian.
In summary, to simulate a given galaxy survey, we need $b(z)$, describing how
clustered its sources are relative to dark matter; $dn/dz$, describing how the
observed sources are distributed along the line of sight; and $\bar{n}$, the
average number density of sources per steradian.

For  $\ell> 20$, we  use the Limber approximation to compute $\Cl^{XY}$. This
dramatically reduces the computation time and gives results that are accurate to
within about 1\%~\cite{LoVerde:2008re}. In this approximation, the
cross-correlations become
\be
C_{\ell}^{XY} = \int dz \frac{H(z)\,D^2(z)}{c\,r^2(z)}\,\left[P(k) \,W^X(k,z)\,W^Y(k,z)\right]_{k=k_{\ell}},
\ee
where $k_{\ell} = (\ell+\tfrac{1}{2})/r(z)$ and $H(z)$ is the Hubble parameter.

We developed an independent code to calculate the cross-power spectra
$C_{\ell}^{XY}$ and have extensively tested its accuracy for various
survey redshift ranges against the publicly available {CLASS} code~\cite{Lesgourgues:2011re}.

\subsection{Simulating LSS maps}\label{sec:lssmap_overview}

As we care only about large-angle ($\ell\lesssim 100$) features, we  model the
ISW signal and galaxy number density fluctuations as correlated Gaussian
fields. To simulate them, we  compute the relevant angular auto- and cross-power
$\Cl$'s and then  use  the {\tt synalm} function from
Healpy~\cite{Gorski:2004by} to generate appropriately correlated sets of
spherical harmonic coefficients $\glm$. These components are defined via the spherical
harmonic expansion of the number density of sources in the $i$th LSS map,
\begin{equation}
\left [\frac{\delta N}{N}\right ]^i(\nhat) = \sum_{\ell m}\,\glm^i\,\Ylm(\nhat).
\end{equation}
For each study using simulated maps, we generate 10,000 map realizations. We
use Healpix with {\rm NSIDE}=32 and compute $\Cl$ up to $\ell_{\rm max}=95$,
guided by the relation $\ell_{\rm max}=3({\rm \tt NSIDE})-1$. Unless we state
otherwise, our ISW reconstructions include multipole information down to
$\ell_{\rm min}=2$.

All of our analyses are for full-sky data and our fiducial cosmological model
is $\Lambda$CDM, with parameter values from best-fit Planck 2015, 
$\{\Omega_ch^2,\Omega_bh^2,\Omega_{\nu}h^2,h,n_s\}=\{0.1188,0.0223,0,0.6774,0.9667\}$.

\subsubsection{Fiducial survey}\label{sec:fidmap}

We model our fiducial galaxy survey on what is expected for
Euclid~\cite{Laureijs:2011gra}. 
With its large sky coverage and deep redshift distribution
the Euclid survey has been identified as a promising tool for ISW
detection~\cite{Afshordi:2004kz,Douspis:2008xv} and it is reasonable to assume that these
properties will also make it a good data set to use for ISW reconstruction.
 We therefore adopt the redshift distribution
used in Ref.~\cite{Martinet:2015wza}, 
\be\label{eq:fiddndz}
\frac{dn}{dz} = \frac{3}{2z_0^2}z^2\,\exp{\left[-(z/z_0)^{-1.5}\right]}
\ee
which has a maximum at $z_{\rm peak}\simeq 1.21z_0$.  We adopt $z_0=0.7$ and
$\bar{n} = 1\times 10^9$. For binning studies (see Sec.~\ref{sec:bintest})
we assume a photo-$z$ redshift uncertainty of $\sigma(z)=0.05(1+z)$. Our
fiducial bias is $b(z)=1$. We explicitly state below whenever these fiducial
values are varied for our tests.

\subsection{ISW estimation}\label{sec:iswest}

We use the optimal estimator derived in Ref.~\cite{Manzotti:2014kta} to reconstruct
 the ISW signal from LSS maps. Because we are interested in quantifying the impact of galaxy survey systematics, in this
work we focus on the case where only galaxy maps are used as
input.  We thus neglect the part of the estimator that includes CMB temperature information and write
\be\label{eq:iswest_simple}
\hat{a}_{\ell m}^{\rm ISW} = \sum_i^nR_{\ell}^i\glm^i.
\ee
Here $\hat{a}_{\ell m}$ is the optimal estimator for the ISW map component,
$\glm^i$ is the observed spherical component of LSS tracer $i$, and $n$ is the
number of LSS tracers considered. The operator

\begin{equation}
  R_{\ell}^i\equiv -N_{\ell}[D_{\ell}^{-1}]_{{\rm ISW-}i}
  \label{eq:R_def}
\end{equation}
is the reconstruction filter  applied to the  $i$th LSS map.  It is constructed
from the covariance matrix $D_{\ell}$ between ISW and LSS tracers,


\be
D_{\ell} = \left(\begin{array}{cccc}
  C_{\ell}^{\rm{ISW,ISW}}&C_{\ell}^{\rm{LSS}_1\rm{,ISW}} &\cdots &C_{\ell}^{\rm{LSS}_n\rm{,ISW}}\\
  C_{\ell}^{\rm{LSS}_1\rm{,ISW}}&C_{\ell}^{\rm{LSS}_1\rm{,LSS}_1} &\cdots &C_{\ell}^{\rm{LSS}_1\rm{,LSS}_n}\\
  \vdots&\vdots &\ddots &\vdots\\
  C_{\ell}^{\rm{LSS}_n\rm{,ISW}}&C_{\ell}^{\rm{LSS}_1\rm{,LSS}_n}&\cdots &C_{\ell}^{\rm{LSS}_n\rm{,LSS}_n}\\
\end{array}\right). 
\ee
The term $N_{\ell}^{-1} \equiv (D_{\ell}^{-1})_{11}$ estimates the reconstruction variance.


Note that for reconstruction using a single LSS map this reduces to a
Wiener filter.
\be
\hat{a}_{\ell m}^{\rm{ISW}} \myeq \frac{\Cl^{\rm ISW-gal}}{\Cl^{\rm gal-gal}}\glm.
\ee
%


In the subsequent discussion, we will refer to the correlations appearing in
$D_{\ell}$ (and thus the reconstruction filters $R_{\ell}^i$) as $\Cl^{\rm
  model}$.  This is to distinguish them from the correlations used to generate
the simulations, which we will call $\Cl^{\rm true}$. We adopt this convention
because if we were reconstructing the ISW signal based on real data, $\Cl^{\rm
  true}$ would be the correlations determined by the true underlying physics
of the universe, while $\Cl^{\rm model}$ would be computed theoretically based
on our best knowledge of cosmological parameters and the properties of the
input LSS tracers.

Setting $\Cl^{\rm model}=\Cl^{\rm true}$ represents a best-case scenario where
we have perfect knowledge of the physics going into the calculations outlined
in Sec.~\ref{sec:clxy}.  Incorrect modeling will break that equality,
causing the estimator in Eq.~(\ref{eq:iswest_simple}) to become suboptimal.
Our analysis of LSS in Sec.~\ref{sec:surveysys} systematics will
fundamentally be an examination of how different manifestations of this kind
of $\Cl^{\rm model}\neq\Cl^{\rm true}$ mismatch impact reconstruction.

\subsection{Fitting for effective galaxy bias}\label{sec:biasfitting}

Our pipeline actually contains an additional step, which as we will see
in later sections, helps protect against some systematics; before
constructing the ISW estimator, we fit the galaxy maps for a
constant bias.

When performing this procedure, the first step of our
reconstruction process is to measure the galaxy
autopower spectrum from the observed galaxy map,  $\Cl^{\rm gal (obs)}$. This will
be subject to cosmic variance scatter about $\Cl^{\rm gal (true)}$ and so will be
realization dependent. We then perform a linear
fit for a constant $\bar{b}$ satisfying
\be
C_{\ell}^{\rm gal (obs)}=\bar{b}^2\, \Cl^{\rm gal(model)}.
\ee
We then scale the model power spectra:
\begin{align}
  \Cl^{\rm gal}\quad&\rightarrow \quad\bar{b}^2\,\Cl^{\rm gal},\nonumber \\
  \Cl^{\rm gal-ISW}\quad&\rightarrow\quad \bar{b}\,\Cl^{\rm gal-ISW},\\
  \Cl^{{\rm gal}_i-{\rm gal}_j}\quad&\rightarrow\quad \bar{b}^i\,\bar{b}^j\,\Cl^{{\rm gal}_i-{\rm gal}_j}.\nonumber  
\end{align}
If there are no systematics affecting our measurements,
$C_{\ell}^{\rm gal (true)} = \Cl^{\rm gal(model)}$,
so $\bar{b}$ will be close to 1. When a galaxy bias is modeled as
a constant, $b_0$, for each galaxy map, this scaling will exactly correct
for any mismatch between the value used in the simulations and that
in the model used to construct the ISW estimator:
\be
\bar{b} = b_0^{\rm true}/b_0^{\rm model}.
\ee
Outside the case of constant bias, there is not a direct correspondence between
$\bar{b}$ and the paramters of the bias model. (It corresponds to the
ratio between  
weighted averages of $b(z)^{\rm true}$ and $b(z)^{\rm model}$.) However, the
procedure for fitting for and scaling by $\bar{b}$ {\it is} well defined and
makes our estimator robust
against systematics which shift $\Cl$'s by a multiplicative constant, including
mismodeled $b(z)$ and $dn/dz$. We will
demonstrate this in Sec.~\ref{sec:zdisttest}.

\subsection{Evaluating reconstruction accuracy}\label{sec:recstats}
We will use two statistics to quantify the accuracy of reconstructed ISW
maps. Primarily, we will use the correlation coefficient  between the
true ISW signal $T^{\rm ISW}(\nhat)$ and the reconstructed ISW map
 $T^{\rm
  rec}(\nhat)$. For a given realization we compute this as
\be
\rho = \frac{\langle T^{\rm ISW}T^{\rm rec}\rangle_{\rm pix}}{\sigma_{\rm ISW}\sigma_{\rm rec}},\label{eq:rhofrommap}
\ee
where $\langle\rangle_{\rm pix}$ indicates an average over pixels, and $\sigma_X$ is the variance of map $X$.

We can approximate the theoretical expectation value for $\rho$  using the cross-power between maps,
\begin{align}
  \langle\rho\rangle =\frac{\sum_{\ell i}(2\ell+1)R_{\ell}^i\Cl^{{\rm ISW}-i}}{\langle \sigma_{\rm rec} \rangle \langle \sigma_{\rm ISW} \rangle},\label{eq:rhoexp}
\end{align}
where the indices $i$ and $j$ label LSS maps and
\begin{align}
  \langle \sigma_{\rm ISW} \rangle &= \sqrt{ \sum_{\ell}\,(2\ell+1)\,\Cl^{\rm ISW}}\\
  \langle \sigma_{\rm rec} \rangle &=\sqrt{ \sum_{\ell i j}\,(2\ell+1)\,R_{\ell}^iR_{\ell}^j\Cl^{ij}}
\end{align}
are the standard deviations of the temperature maps.In deriving this expression, we assumed $\langle \sigma^{-1}\rangle = \langle \sigma\rangle^{-1}$ and that the various factors in this expression are uncorrelated. We will see later that this is a reasonably accurate approximation to make, as it gives values which are in good agreement with simulation results.

One can see by examining Eqs.~(\ref{eq:rhofrommap}) and~(\ref{eq:rhoexp}) that
$\rho$ is sensitive to the reconstruction of phases but insensitive to
changes in the overall amplitude of the reconstructed ISW map.  Because of this,
though $\rho\rightarrow 1$ is generally indicative of a more accurate
reconstruction, this quantity does not capture all important information about
reconstruction quality. We therefore also consider a complementary statistic
which is sensitive to amplitude, defined 
\begin{equation}
s = \frac{\langle (T^{\rm ISW} - T^{\rm rec})^2\rangle_{\rm pix}^{1/2}}{\sigma_{\rm ISW}}.
\end{equation}
The quantity $s$ measures how the average size of errors in the 
reconstructed signal
compares to that of fluctuations in the true ISW map. As with $\rho$, we can
compute its expectation value,
\begin{equation}\label{eq:s_exp}
\langle s\rangle = \frac{\sqrt{\langle \sigma_{\rm rec} \rangle^2 +\langle \sigma_{\rm ISW} \rangle^2   -  2\,\sum_{\ell i}\,(2\ell+1)\, R_{\ell}^i\Cl^{{\rm ISW}-i}}}{ \langle \sigma_{\rm ISW} \rangle}.
\end{equation}

Because the bias-fitting procedure discussed in Sec.~\ref{sec:biasfitting}
corrects for amplitude differences, for most of the scenarios we
study, $\rho$ and $s$ effectively contain the same information. For this
reason, we will primarily
use $\rho$ as our quality statistic and will only show results for $s$ 
when it contributes new insight. 

Throughout this paper we will use angled brackets to
indicate the theoretical expectation values for these statistics,
and an overbar to indicate averages computed from simulations. 


\section{Results I: The effect of survey properties}\label{sec:surveyprops}

Before studying the effects of systematics, it is instructive to explore how
LSS survey properties impact ISW signal reconstruction in the ideal, $\Cl^{\rm
  model}=\Cl^{\rm true}$,  scenario. This has already been done to some
  extent in Refs.~\cite{Manzotti:2014kta},~\cite{Ade:2015dva}, 
and~\cite{Bonavera:2016hbm}.  

Our studies in this section will serve two primary purposes.  First, they will
provide a straightforward demonstration of our pipeline and the reconstruction
quality statistics introduced in Sec.~\ref{sec:recstats}. More importantly,
they will serve as a baseline for our analysis of systematics in
Sec.~\ref{sec:surveysys}:  Our goal is {\it not} to find optimized survey
 properties for ISW signal
reconstruction, though our results might serve as a rough guide for doing so.
Rather, we want to study how shifting, for example,
survey depth or redshift binning strategy affects ISW reconstruction in the
best-case scenario (with no systematic errors) so that we can better
understand the impact of what happens when those errors are introduced.

\subsection{Varying survey depth}\label{sec:depthtest}

\begin{figure}[t]
 \includegraphics[width=1.1\linewidth]{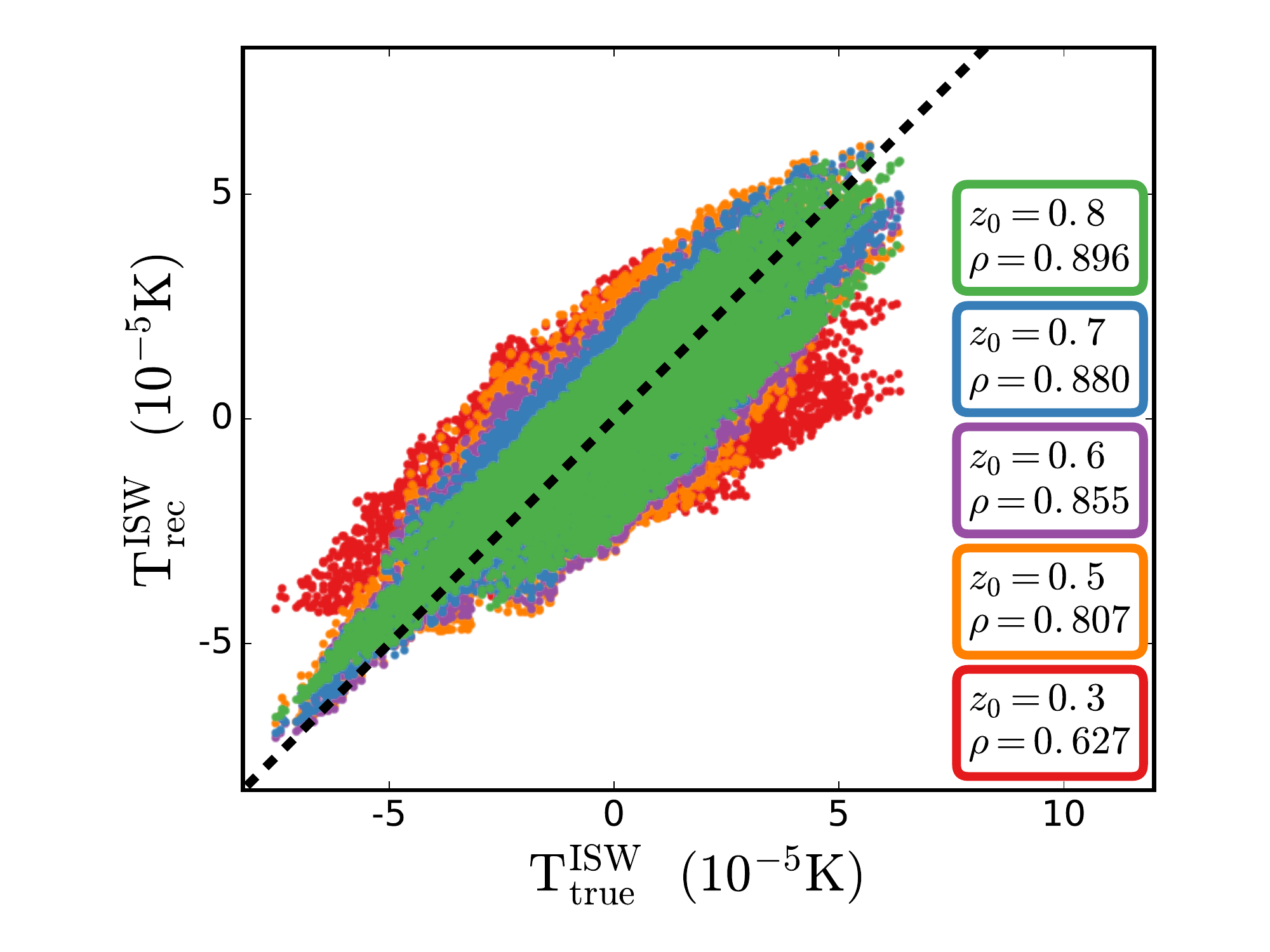} 
  \caption{Scatter plot comparing the true (simulated) ISW signal, on the
    horizontal axis, to the reconstructed ISW signal, on the vertical axis,
    for a single realization assuming  each  of five different depths of
      the survey. Each data point corresponds to one pixel on an {\tt
      NSIDE}=32 map. If there was a perfect reconstruction, all points would
    fall on the dotted gray line.
  }
\label{fig:depthest_TTscatter}
\end{figure}

The first property we examine is survey depth. We model this by changing the
value of $z_0$ in our fiducial $dn/dz$ [Eq.~(\ref{eq:fiddndz})]
 while holding all other survey
properties fixed. We look at values $\Delta z=\pm 0.1$ on either side 
of our fiducial $z_0=0.7$, plus a redshift
distribution comparable to DES~\cite{Becker:2015ilr} with $z_0=0.5$ and the
even-shallower $z_0=0.3$. 

\begin{figure*}[t]
  \includegraphics[width=.49\linewidth]{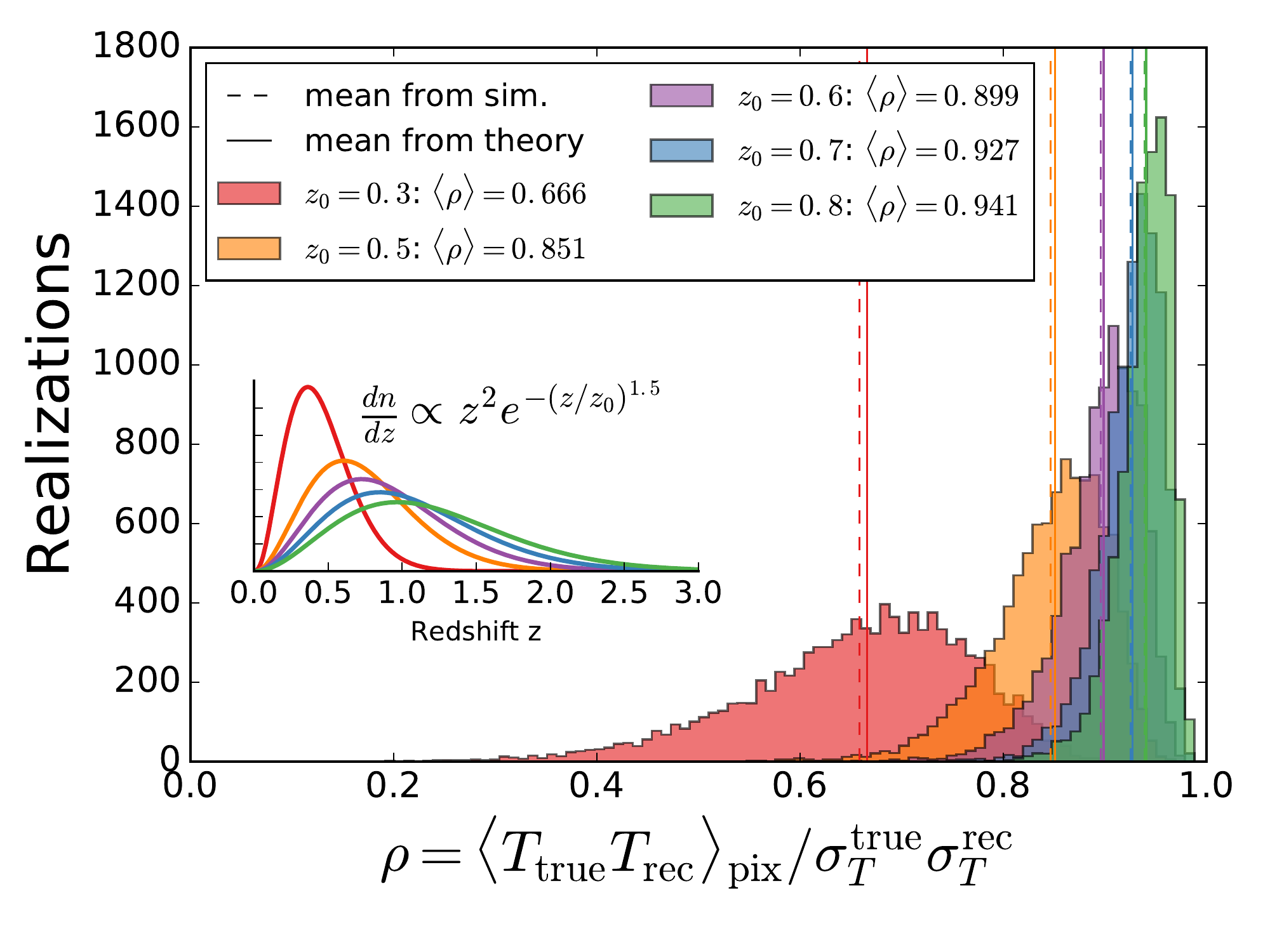}
  \includegraphics[width=.49\linewidth]{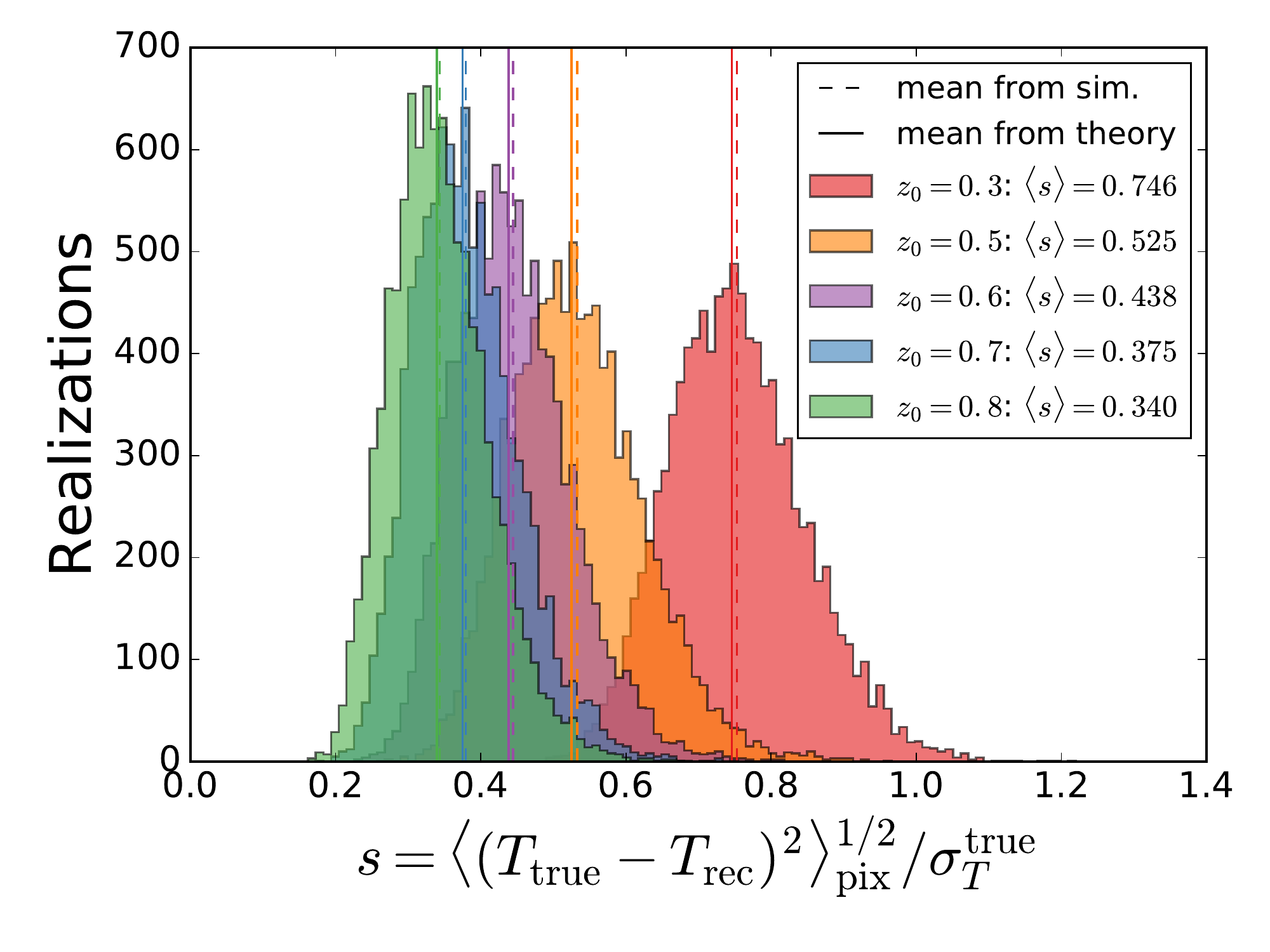}
  \caption{Histograms of the correlation between true and reconstructed ISW
    maps $\rho$ (left panel), and the typical size of residuals relative to
    that of the true ISW map fluctuations $s$ (right panel). These plots
    show the results of 10,000  simulations for surveys of various
    depths, with their $dn/dz$ distributions shown in arbitrary units as an 
    inset in the left plot. The solid and dashed
    vertical lines show the theoretical expectation value and measured
    average, respectively, for the statistic in question.}
  \label{fig:depthtest_hists}
\end{figure*}

Figure~\ref{fig:depthest_TTscatter} shows a pixel-by-pixel comparison between
the reconstructed and true ISW signal for a single representative realization.
We can see that the deeper surveys have data-points more tightly clustered around the
$T^{\rm ISW}_{\rm rec}=T^{\rm ISW}_{\rm true}$ diagonal and correspondingly  higher
values of $\rho$.

We find that this pattern holds, if noisily, in the full ensemble of simulated
maps. Figure~\ref{fig:depthtest_hists} shows histograms of $\rho$ for the same
surveys, with their $dn/dz$ distributions shown in an inset.  In it, the
sample average $\bar{\rho}$ and theoretical expectation value
$\langle\rho\rangle$ are plotted as dashed and solid vertical lines,
respectively. We find that though $\langle\rho\rangle$ tends to be lower than
$\bar{\rho}$, the difference between them is much smaller than the scatter in
the data, and that the ordering of {$\langle\rho\rangle$ values for the
  different surveys is consistent with the results from simulations.  We take
  this to mean that the more computationally efficient $\langle\rho\rangle$ is
  a slightly biased but reasonably reliable indicator of the ISW
  reconstruction quality.

Looking at the data, we also note that the scatter in the individual $\rho$
distributions is large compared to the difference between their mean
values. This tells us that, while $\langle\rho\rangle$ (or $\bar{\rho}$) values
succeed in
predicting how ISW reconstruction quality from different surveys will compare
on average, they are a relatively poor predictor of how surveys will compare for any
individual realization.

For illustrative purposes, in Fig.~\ref{fig:depthtest_hists}, we also show a
histogram for the values of statistic $s$ which, recall, is mainly
sensitive to the amplitude accuracy in the map reconstruction -- measured
from the same simulations. We see that (as expected) surveys with larger
$\bar{\rho}$ have smaller $\bar{s}$ and that the surveys with
$\bar{\rho}\sim0.9$ correspond to $\bar{s}\sim 0.4$. This tells us that even in the best maps
that we study here,  errors in the reconstructed ISW
temperature  are a little over one-third of the amplitude of
true ISW signal fluctuations.

We keep the mean source number density $\bar{n}$ fixed for this analysis, so
that any differences we observe in reconstruction quality are due only to how
the redshift distributions are sampled, not to the fact that a deeper survey will
observe a larger number of sources. We argue that this is well motivated
because the only way $\bar{n}$ enters our calculations is via shot noise,
and we have set it to a large enough
value so that its contributions are negligible on large, ISW-relevant scales.


\subsection{Redshift binning strategy}\label{sec:bintest}
Here we study how different strategies for binning galaxy data affect the
reconstruction. For each bin with $z_i\leq z<z_{i+1}$, we model the redshift
 distribution by weighting the survey's overall
distribution $dn^{\rm tot}/dz$ with a window function $F_i(z)$ and scale the
total number density accordingly:
\begin{align}
  \frac{dn^i}{dz}&=\frac{\displaystyle\frac{dn^{\rm tot}}{dz}\,F_i(z)}
       {\displaystyle\int_{0}^{\infty}\displaystyle\frac{dn^{\rm tot}}{dz}\,F_i(z)\,dz}\label{dndzbin},\\[0.2cm]
       \bar{n}^i &= \bar{n}^{\rm tot}\times
       \left[\int_{0}^{\infty}\displaystyle\frac{dn^{\rm tot}}{dz}\,F_i(z)\,dz\right]\label{nbarbin}.
\end{align}
 We can then compute $\Cl^{XY}$ using
the expressions in Sec.~\ref{sec:clxy}, treating each redshift bin as an
individual map ($X$ or $Y$).

Photometric redshift uncertainties will cause sharp divisions in observed
redshift to be smoothed when translated to spectroscopic
redshift. As in Ref.~\cite{Manzotti:2014kta} we
therefore model the effect of photometric uncertainties $\sigma(z)$ via
\be\label{eq:smoothbinedges}
F_i(z)= \frac{1}{2}\left[{\rm
    erfc}\,\left(\frac{z_i-z}{\sigma(z)\sqrt{2}}\right)-{\rm erfc}\,
  \left(\frac{z_{i+1}-z}{\sigma(z)\sqrt{2}}\right)\right],
\ee
which effectively acts as a smoothed top-hat window in $z$.
We use the standard form for photometric-redshift uncertainty 
\begin{equation}
\sigma(z) = \sigma_{z0}\times (1+z).
\end{equation}
For reference, Euclid forecasts consider $\sigma_{z0}=0.05$  a requirement
and give $\sigma_{z0}=0.03$ as a reach goal \cite{Cimatti:2009is,Laureijs:2011gra}.  

In order to understand how binning affects ISW reconstruction, we split our fiducial redshift distribution into the six bins shown in Fig.~\ref{fig:bintest_zbins} and compute all possible auto- and cross-correlations between them. We then use the relations from Ref.~\cite{Hu:1999ek} to compute $\Cl^{XY}$ for cases where two or more adjacent bins are merged.

\begin{figure}[t]
  \includegraphics[width=1.\linewidth]{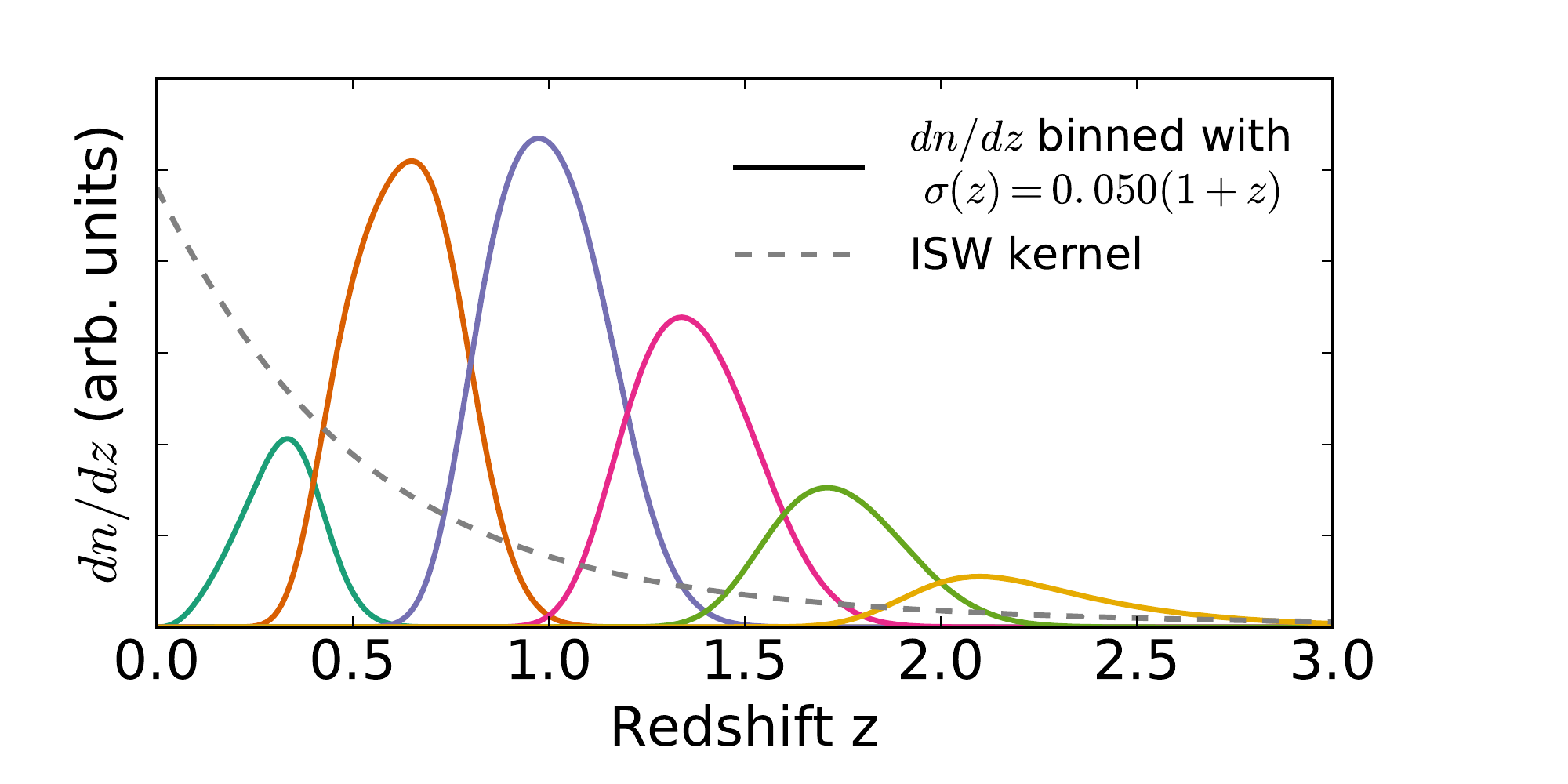}
  \caption{Un-normalized redshift distributions for the six redshift bins
    studied, with photometric-redshift uncertainty $\sigma(z)=0.05
    (1+z)$. Because these distributions are not yet normalized (they neglect
    the denominator of Eq.~(\ref{dndzbin})), the area under the curves
    gives an idea of the relative number of galaxies in each bin. The dotted
    line shows the ISW kernel in arbitrary units. }
  \label{fig:bintest_zbins}
\end{figure}

To check that our understanding of reconstruction statistics  holds for surveys with multiple redshift bins, we simulated 10,000 map realizations for three configurations: the one-bin fiducial case, the six-bin case, and a three-bin case with edges at $z\in[0, 0.8, 1.6,3.5]$. For all of these, we used $\sigma_{z0}=0.05$. The results, shown in Fig.~\ref{fig:bintest_rhohist}, reveal that though binning slightly improves the reconstruction quality, it does not dramatically change the shape of the $\rho$ distribution, nor the relationship between $\langle\rho\rangle$ and $\bar{\rho}$.

\begin{figure}[t]
   \includegraphics[width=1.0\linewidth]{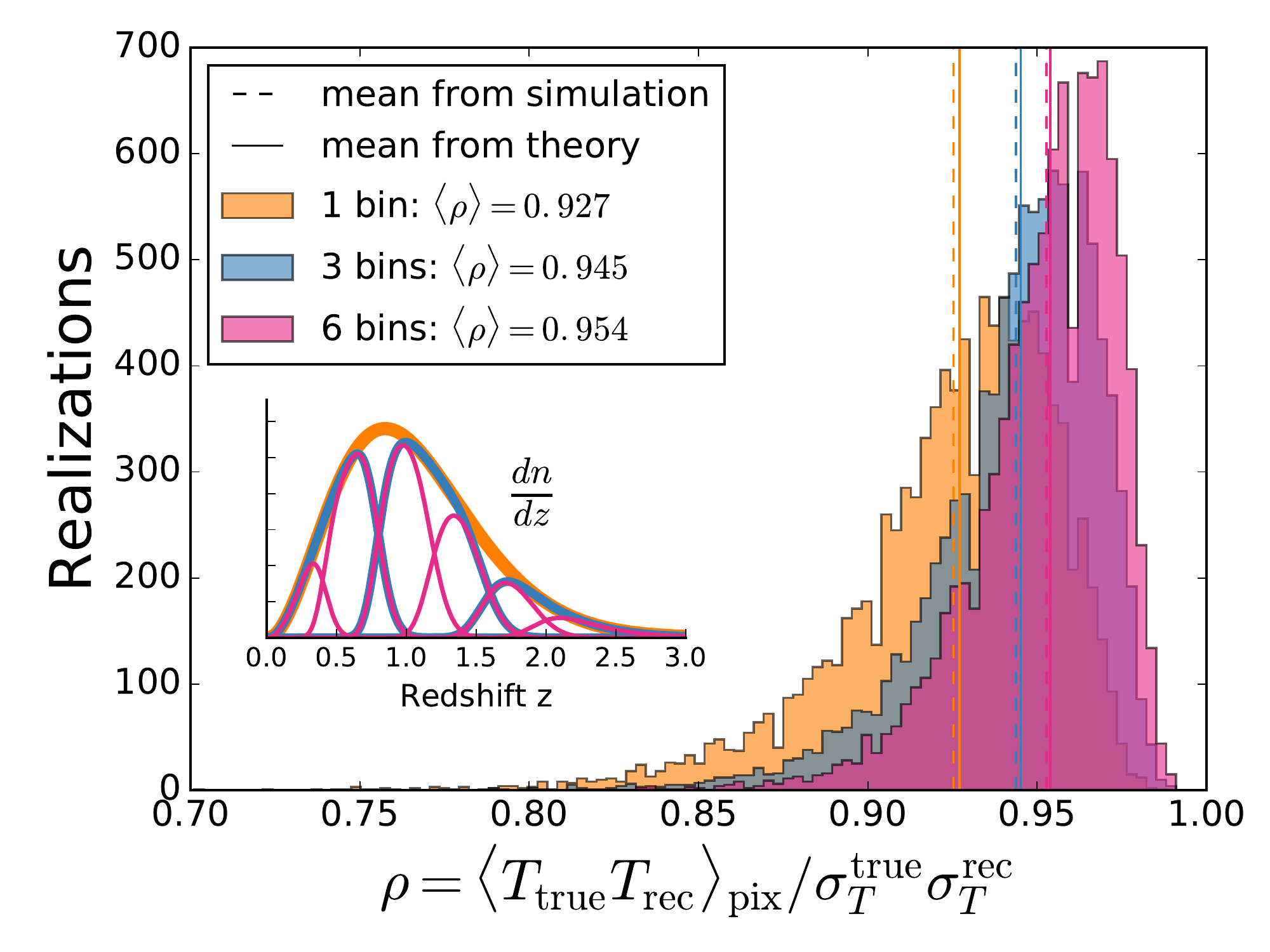} 
  \caption{Histogram of $\rho$ values measured from 10,000 map realizations for selected binning strategies. The inset shows the un-normalized $dn/dz$ distributions for the sets of redshift bins considered.}
\label{fig:bintest_rhohist}
\end{figure}

We see that 
splitting data into
redshift bins improves our ISW reconstruction, if only slightly: the
correlation between the reconstructed and true map shifts by
$\Delta\rho\lesssim 0.03$. This change is smaller than the observed scatter in
$\rho$ and is comparable to that produced in the previous section by
shifting the survey depth by  $\Delta z=\pm 0.1$ about $z_0=0.7$. This improvement could be due to gains in three-dimensional information, or to the fact that we are now using multiple LSS maps with uncorrelated noise.

Reassured that $\langle\rho\rangle$ is still a reliable statistic, we compute it for all 32 possible combinations
of the six bins from Fig.~\ref{fig:bintest_zbins}. The results are shown in Fig.~\ref{fig:bintest_rhoexp}.
In this figure, the bars labeling the $y$-axis schematically illustrate the binning
configurations, with different colors corresponding to different numbers of
bins. The data points
show $\langle\rho\rangle$ for various values of  $\sigma_{z0}$,
while the X-shaped points with error bars show the mean and standard
deviations extracted from the histograms in Fig.~\ref{fig:bintest_zbins}.

\begin{figure*}
 \includegraphics[width=0.8\linewidth]{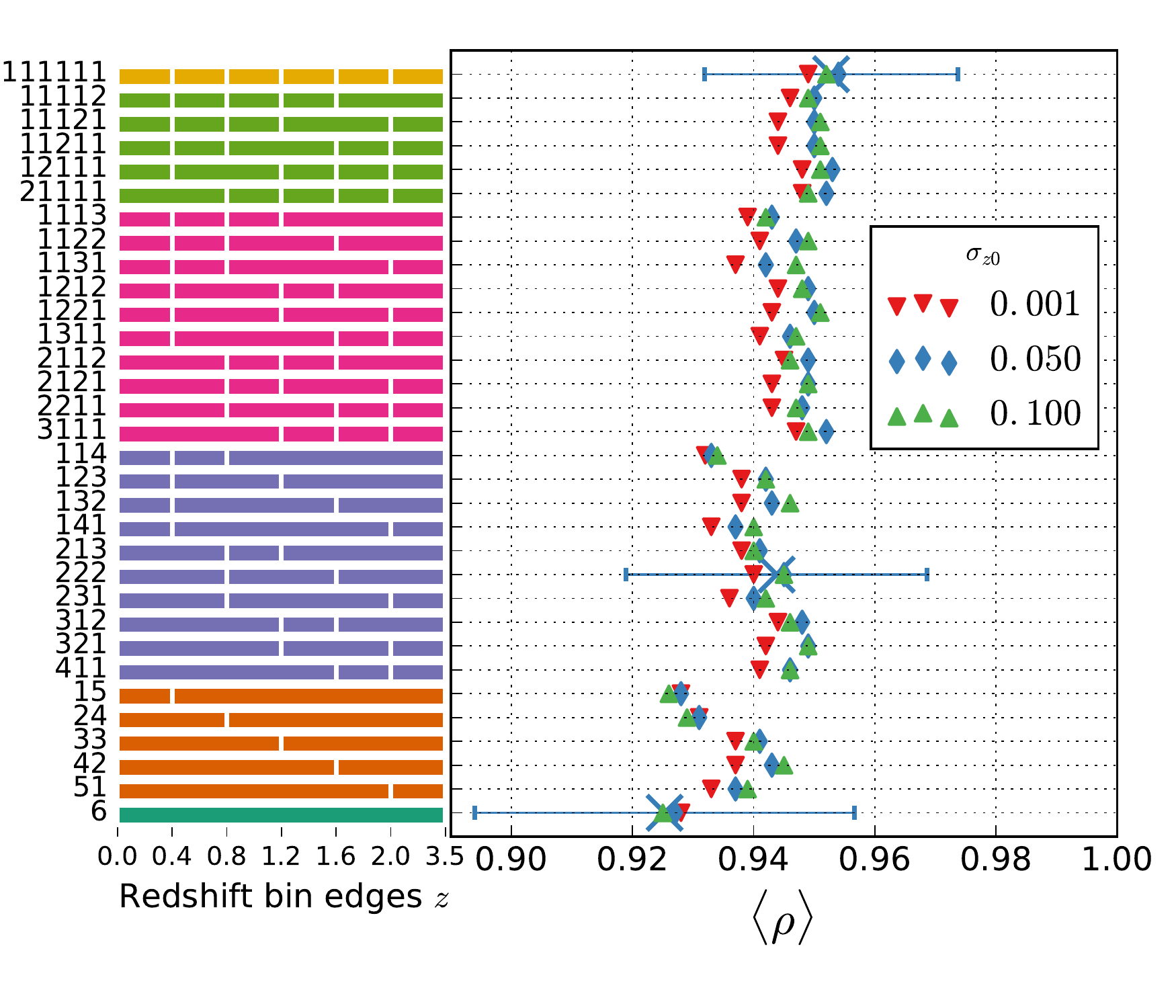} 
  \caption{Theoretical expectation value for $\rho$ computed for different
    redshift binning strategies and levels of photometric-redshift
    uncertainties. The colored bars and corresponding numbers on the
    left side of the plot are schematic labels for how the galaxies are
    divided into redshift bins. Different colored points show the effect of
    different photo-$z$ uncertainties. The ``X'' points with blue horizontal errorbars show
    the mean and standard deviation of $\rho$ extracted from the histograms in
    Fig.~\ref{fig:bintest_rhohist}.}
  \label{fig:bintest_rhoexp}
\end{figure*}

We note a couple of patterns in the results.  First, for a fixed number of
bins, the reconstruction tends to be better if we place finer divisions at
high redshift. Also, having a smaller photometric-redshift uncertainty
actually slightly degrades the reconstruction rather than improving it. This
implies that combining maps with redshift distributions which overlap more
tend to lead to better reconstructions. This could be due a multitracer
effect, in that overlap between bins means that we are sampling the same
potential fluctuations with multiple source populations.  However, it is also
possible this is due to how our model of $\sigma(z)$ affects the shapes of the
redshift distributions.  Given the small size of these effects,  one
should be cautious about assigning them much physical significance.

Last, we observe a shift $\Delta\rho$ due to changes in binning that is
smaller than what is found in the work by\citet{Manzotti:2014kta} by about a factor of
3. Because their simulated DES-like survey is shallower than our fiducial
survey and the relationship between $\Delta\rho$ and $\bar{\rho}$ is
nonlinear (e.g., a shift from 0.98 to 0.99 is more significant than one from
0.28 to 0.29), this does not necessarily mean that our results are
incompatible. As a cross-check, we performed additional simulations similar to
those analyzed in Ref.~\cite{Manzotti:2014kta}. Our results, discussed in
Appendix~\ref{app:MDcrosscheck}, support this.

\subsection{Varying $\ell_{\rm min}$ of reconstruction}\label{sec:lmintest}

For most of the studies presented in this paper, we reconstruct and
assess the accuracy of ISW maps using all multipoles with $2\leq \ell \leq95$.
This range is
chosen because $\ell=2$ is the lowest multipole typically considered for CMB
analysis and $\ell=95$ is the maximum multipole retaining information in 
{\tt NSIDE}=32 Healpix maps. In this section, we study the effect
of changing $\ell_{\rm min}$. 

When we perform  ISW map reconstruction, we enforce $\ell$-range
requirements in three ways. First, when we construct the ISW estimator shown
in Eq.~(\ref{eq:iswest_simple}), we set all $R_{\ell}^i$ not satisfying
$\ell_{\rm min}\leq \ell\leq \ell_{\rm max}$ to be zero, so the
reconstructed map contains no information from multipoles outside that
range. Second, when analyzing simulations, we remove the same $\ell$ values 
from maps before computing $\rho$. Likewise, when
we analytically compute $\langle\rho\rangle$ as shown in
Eq.~(\ref{eq:rhoexp}), we restrict the sum over multipole to $\ell_{\rm
  min}\leq \ell\leq \ell_{\rm max}$. In other words, when we show
$\rho_{\ell\geq \ell_{\rm min}}$, we are showing the result for an ISW map
reconstructed for a limited range of $\ell$ values, evaluated by considering
only those multipoles. 

The results of this analysis are shown in Fig.~\ref{fig:lmintest_fid}. Here
we show the correlation coefficient between true and reconstructed maps
$\rho_{[\ell\geq\ell_{\rm min}]}$ as a function of the minimum multipole used
in the reconstruction. The solid line is the theoretical expectation value,
while the data points with error bars show results from simulations. We find
that $\rho$ increases with the minimum multipole out to $\ell_{\rm min}\sim
5$, after which it begins to very gradually decrease with $\ell_{\rm
  min}$. Increasing $\ell_{\rm min}$ also decreases the scatter in $\rho$
measured across realizations.

We interpret these trends to be the result of a competition between cosmic
variance and the fact that most ISW information (power and cross-power) is at
small multipoles. That is, removing the lowest few multipoles (out
  to $\ell\simeq 4$) from the analysis largely removes noise due to cosmic
  variance, while removing further multipoles largely removes ISW information.
  This has implications for efforts to reconstruct ISW maps
from data; if we only care about small-angle features, it can be worth
ignoring a few low-$\ell$ modes in order to get a more accurate
reconstruction. Conversely, if we want to study how the ISW signal contributes
to the CMB quadrupole and octupole, we must recognize that reconstruction
quality will be necessarily less predictable.

Because cosmic variance of the ISW $\Cl$ has a nontrivial relationship with  the value and scatter of $\rho$, one cannot make a direct connection between
$\ell_{\rm min}$ and how $f_{\rm sky}$ affects reconstruction, as is done in
the ISW signal-to-noise detection studies 
(e.g., Ref.~\cite{Douspis:2008xv}). To understand how sky coverage affects
reconstruction, one should perform simulations using the mask appropriate for a
given survey. We refer the reader to Ref.~\cite{Bonavera:2016hbm} for an analysis
of how ISW signal reconstruction is affected by survey masks. 

We also looked at the impact of varying $\ell_{\rm max}$ but found that
the correlation coefficient $\rho$ is insensitive to it, and therefore
do not show it.

\begin{figure}
 \includegraphics[width=\linewidth]{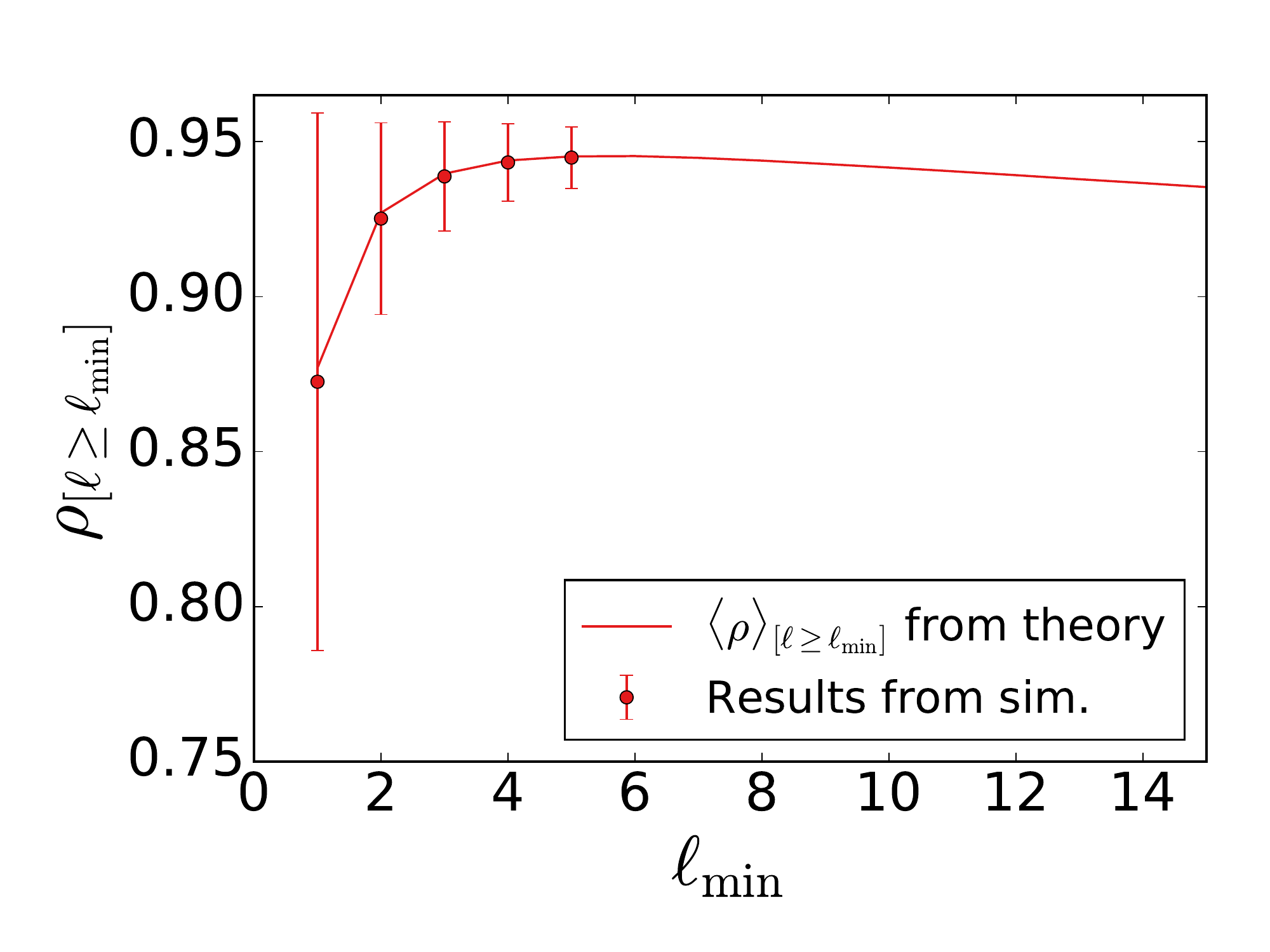} 
 \caption{How filtering out angular scales with $\ell>\ell_{\rm min}$ affects reconstruction of ISW map. The data points show the mean and standard deviation of $\rho$, the correlation coefficient between true and reconstructed ISW maps, observed in 10,000 realizations, while the line shows the value of $\langle\rho\rangle$ computed analytically.}
  \label{fig:lmintest_fid}
\end{figure}

\subsection{Varying $\bar{n}$}\label{sec:shottest}
Additionally, we studied how the level of galaxy shot noise affects
reconstruction.  For this test, we varied the number density of sources,
$\bar{n}$, for our fiducial survey and introduced it to both $\Cl^{\rm true}$
and $\Cl^{\rm model}$ according to Eq.~(\ref{eq:shotnoise}). Our results are
shown in Fig.~\ref{fig:shottest}.

We find that as long as
$\bar{n}\gtrsim 1 \,{\rm arcmin}^{-2}\approx 10^7\,{\rm sr}^{-1}$, shot noise will have a negligible impact on
reconstruction. Note that this requirement is easily satisfied by essentially
all photometric surveys (e.g., for DES or Euclid,
$n\simeq (10-30)\,{\rm  arcmin}^{-2}$). However, the quality of the
reconstruction degrades rapidly
for lower values of number density; once $\bar{n}\lesssim
10^{-3} \,{\rm arcmin}^{-2}\approx 10^4 \,{\rm sr}^{-1}$, the reconstruction
contains effectively no information about the true ISW map.  Therefore, ISW
reconstruction from spectroscopic galaxy surveys, as well as galaxy cluster
samples, may be subject to degradations due to high shot noise. 

\begin{figure}
 \includegraphics[width=\linewidth]{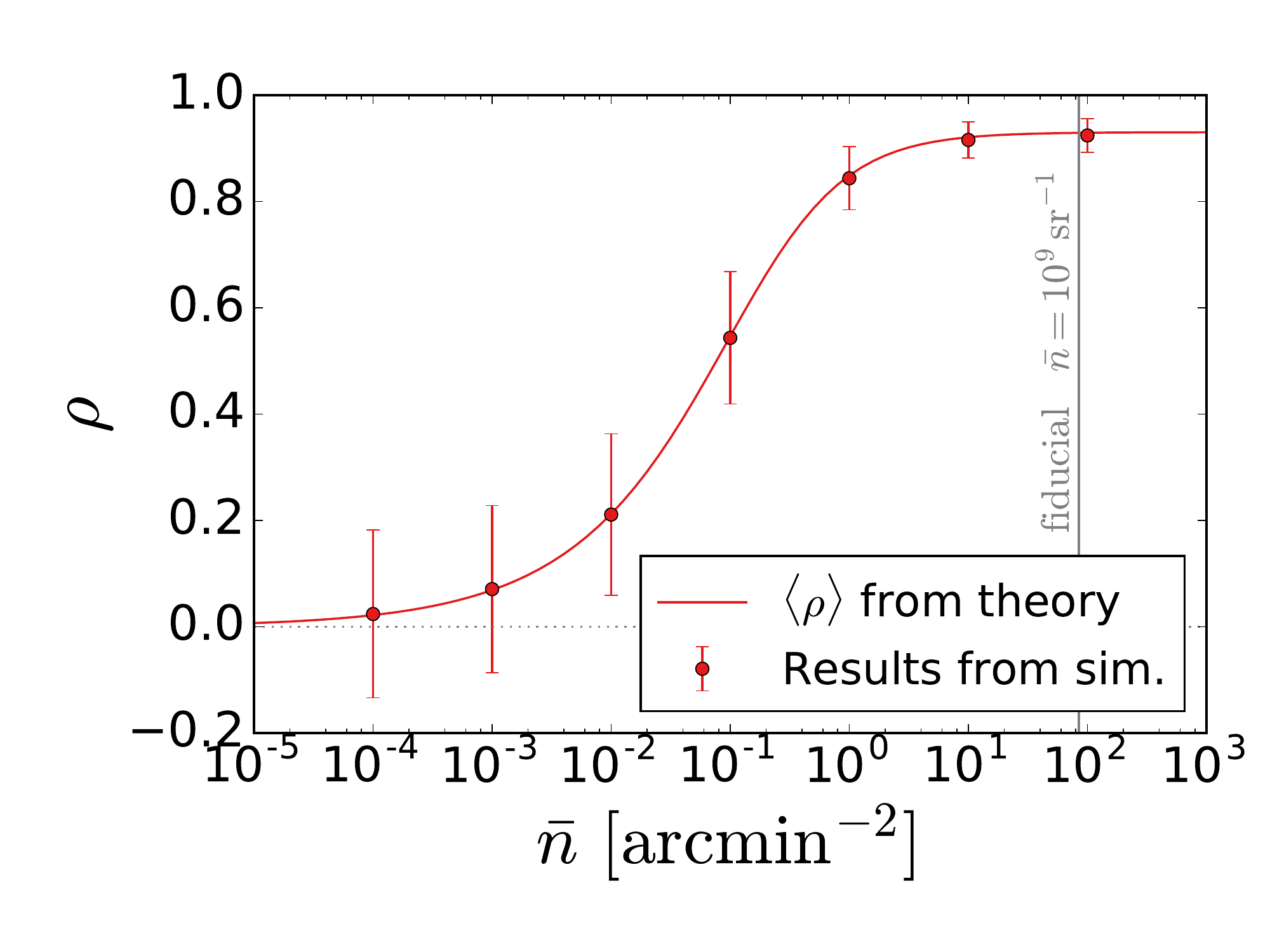} 
 \caption{How changing $\bar{n}$ affects reconstruction of the ISW map. The
   datapoints show the mean and standard deviation of $\rho$, the correlation
   coefficient between true and reconstructed ISW maps, observed in 10,000
   realizations. The line shows the value of $\langle \rho\rangle$ computed
   analytically.}
  \label{fig:shottest}
\end{figure}

\section{Results II: The effect of survey systematics}\label{sec:surveysys}


Large-scale structure surveys are subject to a variety of systematic errors
that limit the extent to which LSS tracers can be used to probe dark matter,
dark energy, and primordial physics.  These systematics can be astrophysical,
instrumental, or theoretical in origin.  Concretely, in this work, they include
anything that makes $\Cl^{\rm model}\neq\Cl^{\rm true}$, which will cause the
estimator given in Eq.~(\ref{eq:iswest_simple}) to become suboptimal.  Our
goal is to study these LSS systematics generally, without requiring specific
information about a LSS survey (e.g., wavelengths at which it observes the
sky).  We do this by considering two broad classes of LSS systematics:
\begin{enumerate}
\item Mismodeling of the distribution of LSS sources along the line of sight.
\item Direction-dependent calibration errors.
\end{enumerate}
Our studies will give us some insight into which, and how much, systematics
need to be controlled if one wishes to use
LSS data to reconstruct a map of the ISW signal.

\begin{figure*}
  \includegraphics[width=.49\linewidth]{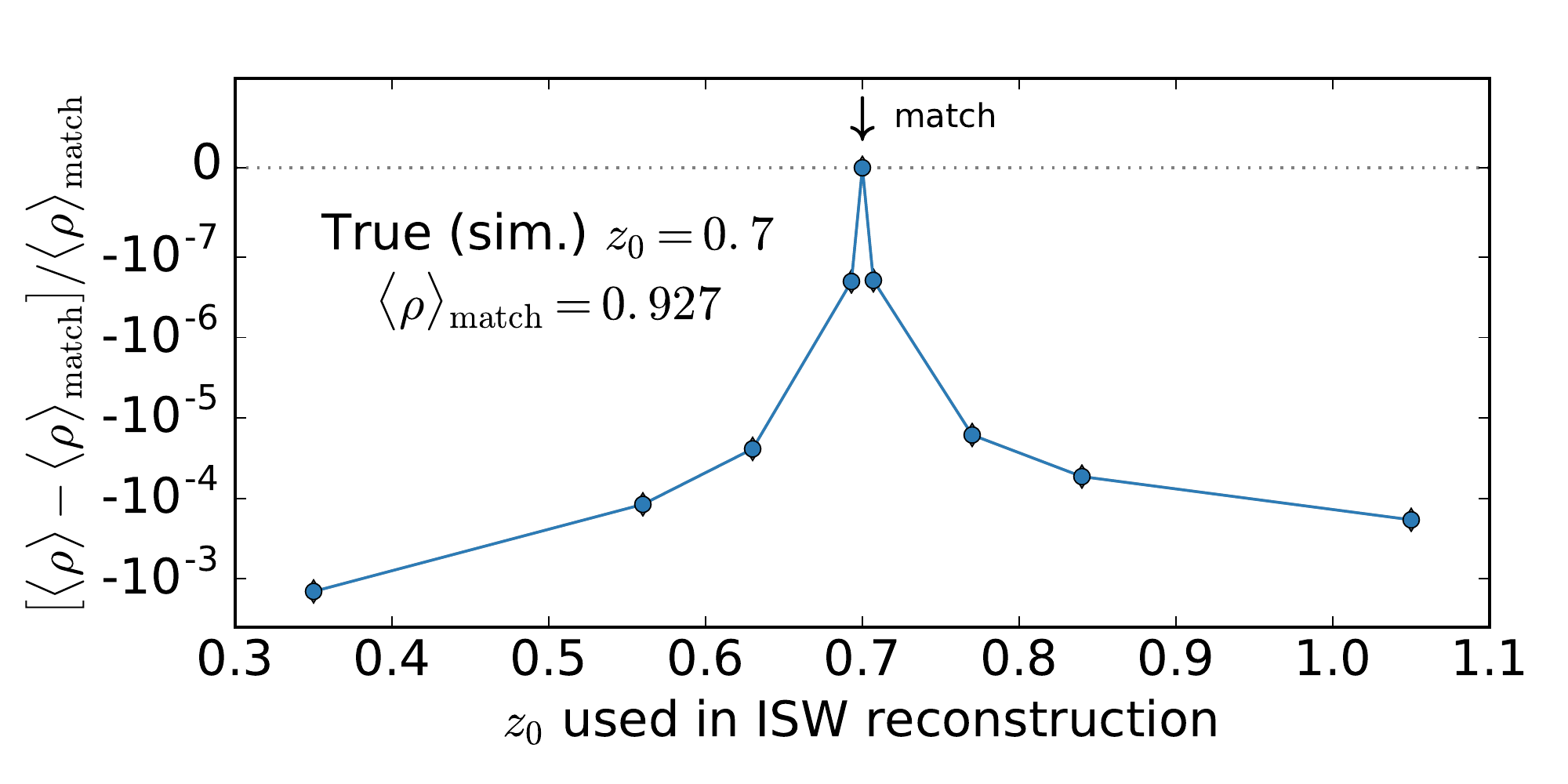}
  \includegraphics[width=.49\linewidth]{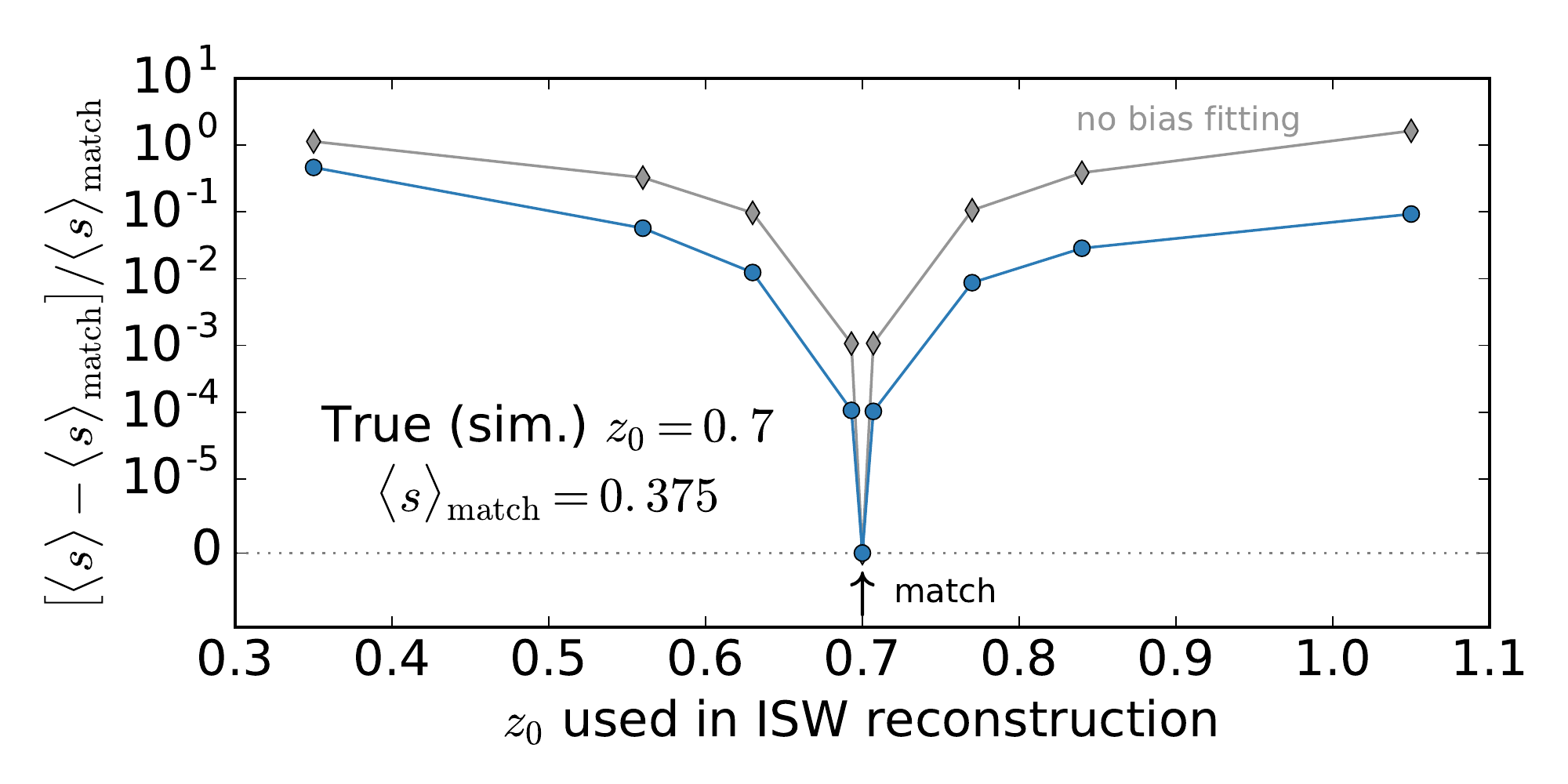}
\caption{Impact of mismodeling survey depth on
  the expected correlation between the true and reconstructed ISW maps
  $\langle \rho\rangle$ (left panel) and the ratio of the average size of residuals 
  to that of ISW map features $\langle s\rangle$ (right panel).
  The   true value of the parameter $z_0$, which controls the depth of
  the survey, is fixed at 0.7, while the values used for reconstruction are
  shown on the $x$-axis.
  The blue circular points show results from our
  standard reconstruction pipeline, while the gray diamond-shaped points (directly
  behind the blue points in the $\rho$ plot) show results when we skip the
  $\bar{b}$-fitting step.  The $y$-axis is linear within one tick mark of zero;
  otherwise, it has logarithmic scaling.
}
  \label{fig:z0test_rho_s_exp}
\end{figure*}

\subsection{Modeling redshift distribution of sources}\label{sec:zdisttest}

In the context of ISW map reconstruction, it would be reasonable to
  guess that accurate knowledge of galaxy redshifts is
important for our ability to correctly associate the observed number density
fluctuations on the sky with the three-dimensional gravitational potential fluctuations
which source the ISW signal. Uncertainties about  redshift distributions are a
pervasive class of systematics affecting LSS surveys, which have already been studied by
numerous authors (e.g., Refs.~\cite{Ma:2005rc,Bernstein:2009bq}) in the context of
cosmological parameter measurements from photometric surveys. Here we study how
redshift modeling errors affect the ISW reconstruction accuracy.

For the purposes of this discussion, we define redshift uncertainties broadly as
anything that makes the galaxy window function (Eq.~(\ref{eq:windowgal})) used in
our ISW estimator different from that which describes the the true line-of-sight
distribution of objects we observe on the sky. We study three specific cases of
this:
the mismodeling of a survey's median redshift, redshift-dependent bias, and the
fraction of catastrophic photometric-redshift errors. In each case, we identify a
parameter which controls the survey characteristic
in question. Then, choosing a true (simulation) value for that parameter, we
perform reconstructions using several mismodeled values as input to the
ISW estimator. This allows us to and look at how the theoretical expectation values
of our quality statistics respond relative the best, correctly modeled case.

Let us place these shifts in context by referring to previous sections.  In an ideal scenario with no systematic errors,
changing the survey depth parameter (see Section~\ref{sec:depthtest}) from the
fiducial $z_0=0.7$ to 0.6 (0.8) causes $\langle\rho\rangle$ to change by 3\%
(1.5\%) and $\langle s\rangle$ by 20\% (10\%). Also,  splitting our fiducial survey
into $6$ redshift bins (in Sec.~\ref{sec:bintest}) improves $\langle\rho\rangle$ by 3\% relative to the one-bin case.

\begin{figure*}
  \includegraphics[width=.49\linewidth]{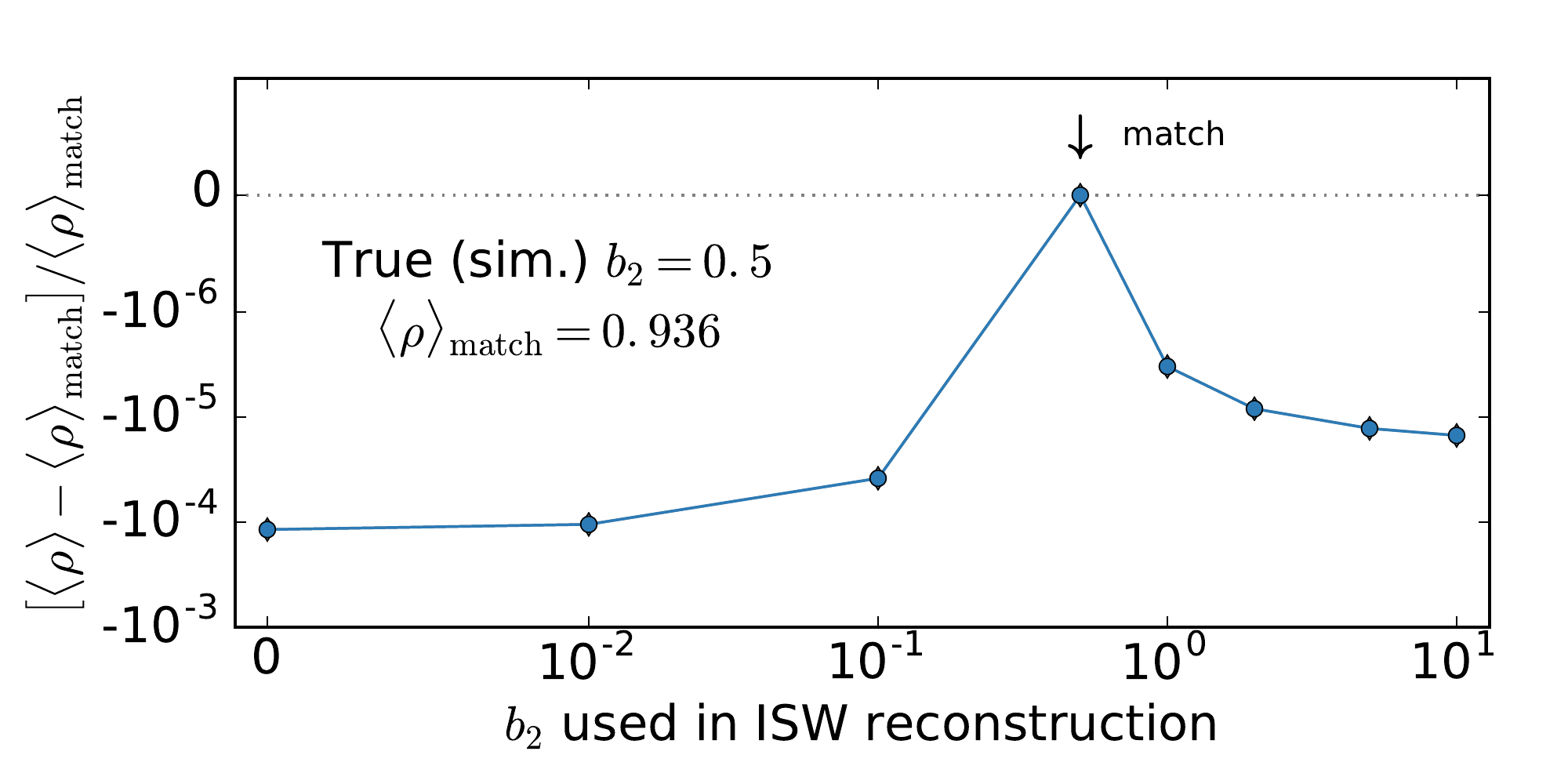}
  \includegraphics[width=.49\linewidth]{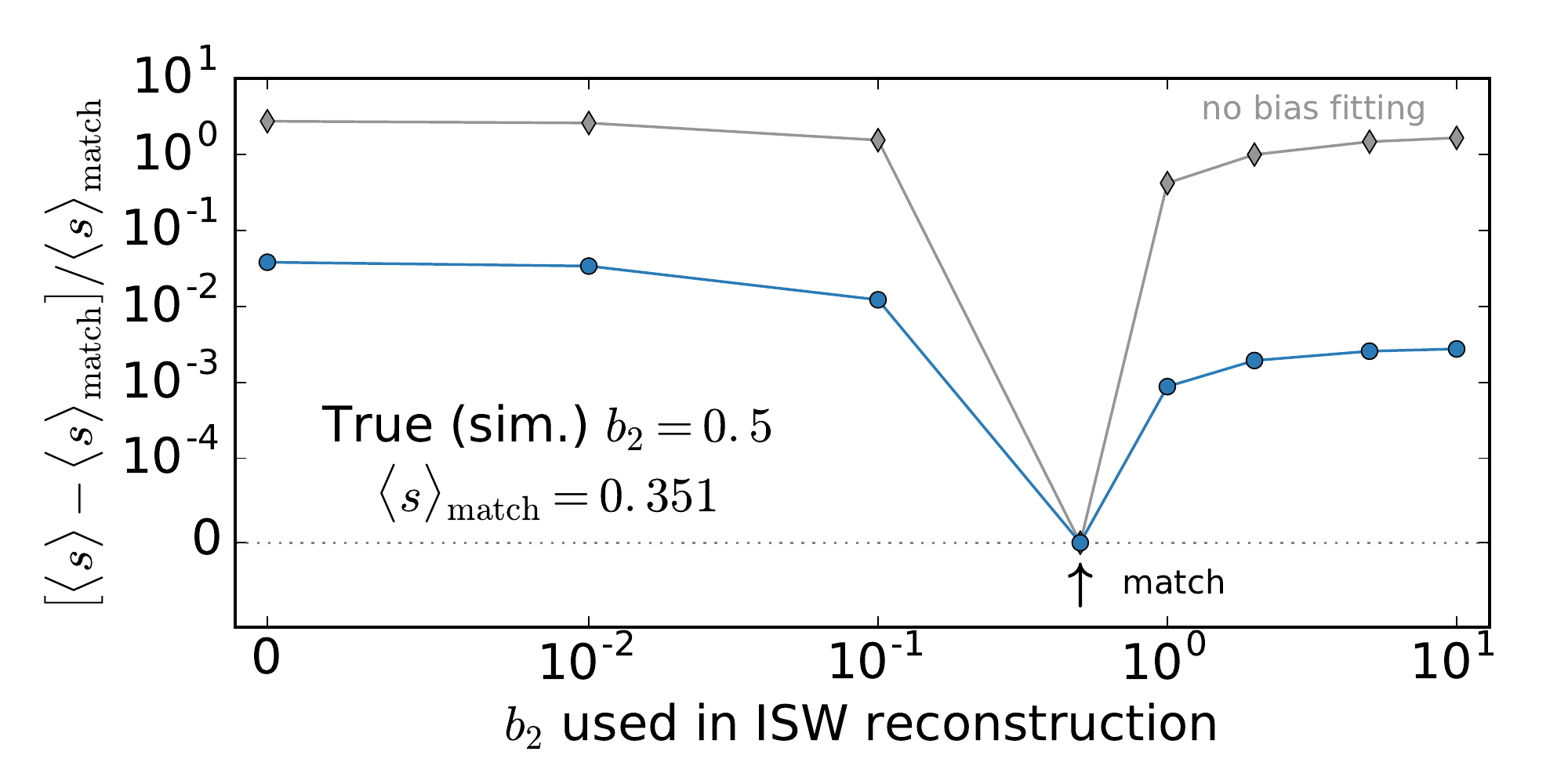}
  \caption{ Impact of mismodeling redshift-dependent bias on the expected
    correlation between the true and reconstructed ISW maps $\langle
    \rho\rangle$ (left panel), and the typical size of residuals relative to
    that of ISW map features $\langle s\rangle$ (right panel). The bias is
    modeled as $b(z)=1+b_2(1+z)^2$ with the true value fixed at $b_2=0.5$ and
    the values used in the ISW estimator shown on the $x$-axis.  Both axes
    have logarithmic scaling except in regions within one tick mark of zero,
    where they are linear. The blue circular points show results from our
    standard reconstruction pipeline, while the gray diamond-shaped points
    (directly behind the blue points in the $\rho$ plot) show results when we
    skip the $\bar{b}$-fitting step.}
  \label{fig:bztest_rho_exp}
\end{figure*}

\subsubsection{Median redshift}\label{sec:z0test}

We begin by studying how reconstruction accuracy responds when we
construct the ISW estimator using the wrong median LSS source redshift.
Though the parameter $z_0$ in the $dn/dz$ distribution given in
Eq.~(\ref{eq:fiddndz})  is lower than $z_{\rm  median}$, raising or lowering it will
have a similar effect as shifting the median of the distribution. We thus
use $z_0$ as a proxy for median redshift. We compute $\Cl^{\rm true}$ with $z_0$ fixed
at its fiducial value of 0.7, and vary the $z_0$ values used to compute
$\Cl^{\rm model}$.

Figure~\ref{fig:z0test_rho_s_exp} shows the fractional change in
our reconstruction statistics
when the value of $z_0$ used for
reconstruction is shifted from its true value by $\pm 1\%$, $\pm 10\%$, $\pm
20\%$, $\pm 30\%$, and $\pm 50\%$. We see that even for large shifts in $z_0$
(with correspondingly dramatic mismatches between the true and
model $dn/dz$) the fractional change in $\rho$ is less than
$\mathcal{O}(10^{-3})$.  The effect on $s$ is also small; for all but the most
extreme points, the fractional change in the size of residuals
$\langle s \rangle$ is less than 10\%.

To understand this lack of sensitivity of $z_0$, it is instructive to note that varying $z_0$
changes $\Cl$ by a nearly scale-independent amplitude. (See
Appendix~\ref{sec:zdisttest_cl_plots} for plots demonstrating this.)
As we observed in Sec.~\ref{sec:recstats}, $\rho$, the correlation coefficient
between true and reconstructed ISW maps, is
insensitive to overall shifts in the the map amplitude. The fact that it does not
respond strongly to these changes in $z_0$ is thus not surprising.
The statistic $\langle s \rangle$, which measures the size of residuals,
is sensitive to changes in
amplitude, however. The fact that it also displays small fractional changes
illustrates the importance of the bias-fitting procedure described in
Sec.~\ref{sec:biasfitting}. Because the effects of mismodeling $z_0$ are 
degenerate with shifts in constant bias, fitting for $\bar{b}$ 
protects our reconstruction against this kind of systematic.

For comparison, we compute $\langle\rho\rangle$ and $\langle s\rangle$ while
neglecting the bias-fitting step and show the results as gray points in
Fig.~\ref{fig:z0test_rho_s_exp}. We see no change in the $\rho$ plot (the
gray points are directly behind the blue ones), reflecting the fact that
$\rho$ is insensitive to constant multipliers. In the $s$ plot, we see that the
bias-fitting procedure suppresses the size of the reconstruction errors by about an
order of magnitude.

To summarize, we find that the quality of the ISW reconstruction is much less
dependent on our knowledge of the survey's median redshift than naively
expected.  The median redshift mostly changes the normalization of the $\Cl$,
but so does the galaxy bias (which, recall, is to a good approximation
scale independent at the large scales we are studying).  By fitting for the bias
parameter in the angular power spectrum---something that is typically done in
LSS surveys regardless of their application---one effectively also fits for
$z_0$. As a result, the combination of the  galaxy bias and survey depth that
enters the amplitude of the $\Cl$ is fit to the correct value.

\subsubsection{Redshift-dependent bias}\label{sec:bztest}

Here, we study what happens if the redshift dependence of the galaxy bias is
modeled incorrectly. Using the functional forms  given
in Ref.~\cite{Ade:2015dva} for guidance, we parametrize
the redshift dependence of the bias via
\begin{equation}\label{eq:bzmodel}
  b(z)=b_0(1+b_2(1+z)^2).
\end{equation}
For this study, we set $b_0=1$ and vary $b_2$, noting that Ref~\cite{Ade:2015dva} uses
$b_2\sim0.5$ for sources in NVSS and WISE-AGN.

In the expression for $\Cl$, $b(z)$ appears inside the same integrand as
$dn/dz$, so changes to $b(z)$ have an effect similar to altering the LSS source
redshift distribution. The results here, shown in
Fig.~\ref{fig:bztest_rho_exp}, are thus similar to what was seen in the
previous section. Increasing $b_2$ mostly just increases the overall amplitude
of the galaxy $\Cl$'s, so the reconstruction is not very sensitive to $b_2$
once we fit for $\bar{b}$. For example, if the true value of $b_2$ is 0.5 and we
reconstruct the ISW signal assuming no redshift dependence ($b_2=0$), the
fractional change in $\langle\rho\rangle$ is $\mathcal{O}(10^{-4})$ and the
fractional change in $\langle s\rangle$ is $\mathcal{O}(10^{-2})$. The reason
the $\bar{b}$-fitting step has a larger effect here than in the $z_0$ study above
is probably because the normalization requirements of $dn/dz$ somewhat limit
the size of $\Cl$ amplitude shifts, whereas $b(z)$ has no such normalization
scaling.

\subsubsection{Catastrophic photo-z error rate}\label{sec:catztest}

\begin{figure*}
  \includegraphics[width=.49\linewidth]{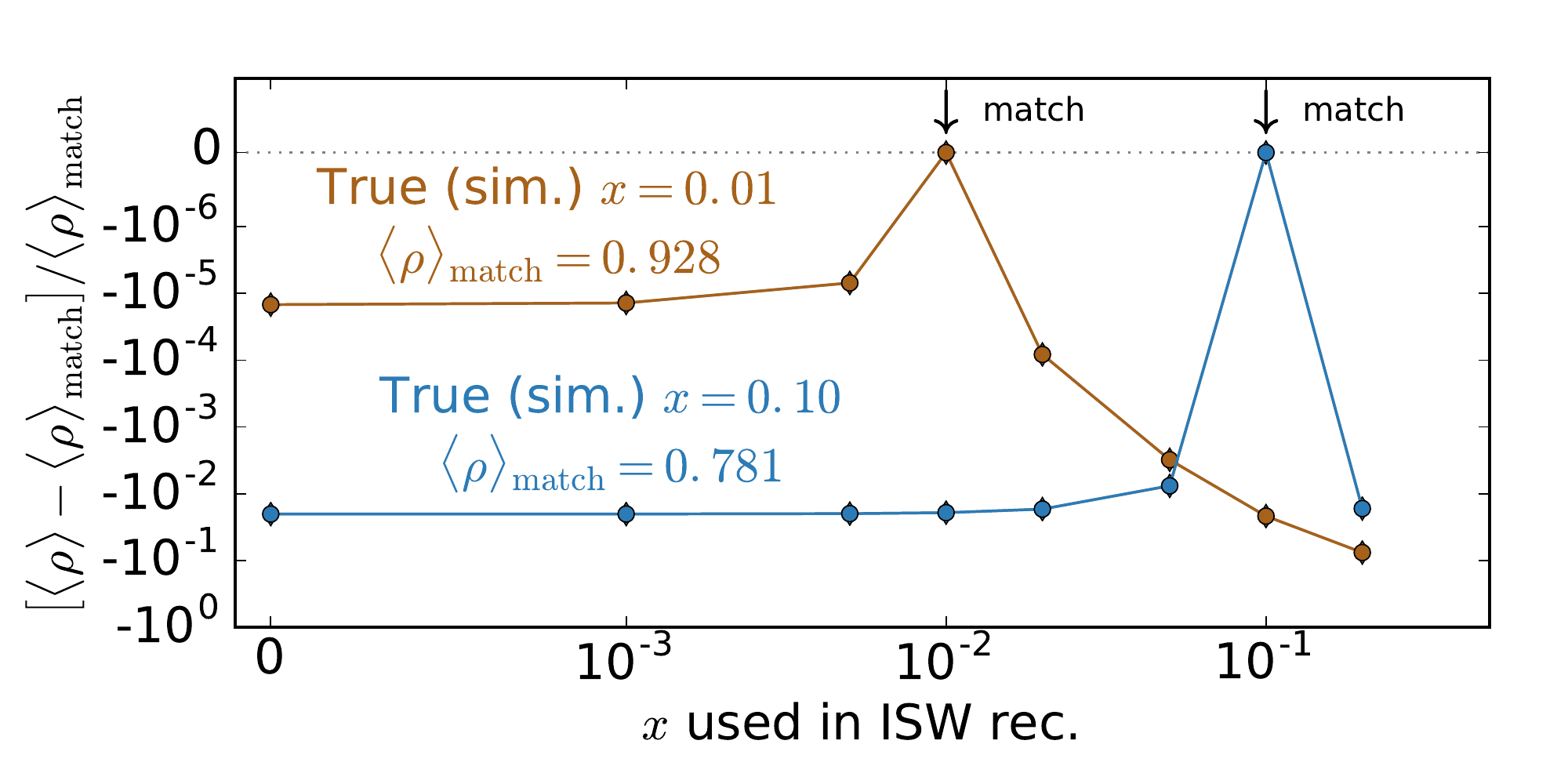}
  \includegraphics[width=.49\linewidth]{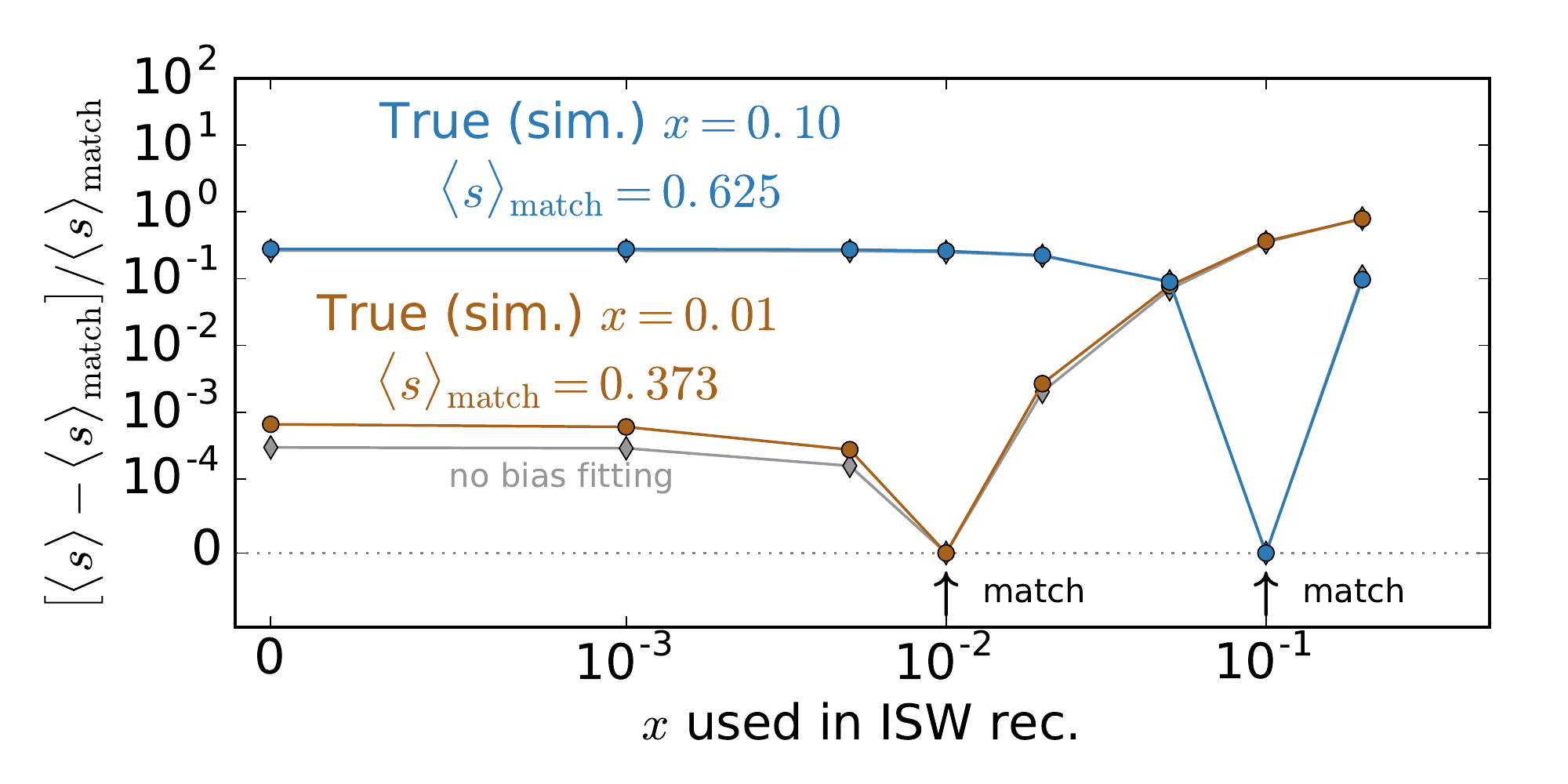}
  \caption{ Impact of mismodeling the fraction $x$ of galaxies subject
    to catastrophic photo-$z$ errors on the expected correlation between the
    true and reconstructed ISW maps $\langle \rho\rangle$ (left panel),
    and the typical size of residuals relative to that of ISW map features
    $\langle s\rangle$ (right panel). 
    Both axes have logarithmic scaling except in regions within
    one tick mark of zero, where they are linear.
    The blue and brown circular
    points show results from our  standard reconstruction pipeline when the true
    value of $x$  is  0.1 and 0.01 respectively. The gray
    diamond-shaped points (directly behind the other points in the $\rho$ plot and
    the blue points in the $s$ plot)
    show results when we skip the $bar{b}$-fitting step.}
  \label{fig:catztest_rho_exp}
\end{figure*}

Galaxies in photometric-redshift surveys are also
  subject to so-called {\it catastrophic} photometric-redshift errors---cases where the true redshift is misestimated by a significant amount
  \cite{Bernstein:2009bq,Hearin:2010jr}.  This is a distinct effect from the
photo-$z$ uncertainty modeled in the binning tests in Sec.~\ref{sec:bintest}, which
causes a redshift bin selected using sharp cuts in photo-$z$ to
occupy a smoothed distribution in the spectroscopic redshift. Rather, for galaxies
suffering catastrophic photo-$z$ errors, the photometric-redshift finding
algorithms have failed, and the spectroscopic redshift corresponding to a given photo-$z$
is effectively randomized. The reasons for this are not fully understood, but, like the
conventional photo-$z$ error case, the rate and outcome of catastrophic errors
depend strongly on the number of photometric filters and their relation to the
spectral features that carry principal information about the redshift.

In the absence of detailed, survey-specific information about the photometric pipeline,
we model catastrophic redshift errors by randomly assigning the true redshift of a
fraction $x$ of the galaxies in our sample (e.g., $x=0.01$ means that one in a hundred
galaxies has a catastrophic photo-$z$ error).  We implement this by modifying the redshift
distribution of each bin $i$ to
\be\label{eq:dndz_withcatz}
\frac{d\tilde{n}^i}{dz}=(1-x)\frac{dn^i}{dz} + x\bar{n}^i\left[\Theta(z-z_{\rm
  min})-\Theta(z_{\rm max}-z)\right],
\ee
where $x$ is the fraction of
galaxies suffering catastrophic errors, $dn^i/dz$ is the redshift distribution
of bin $i$ without catastrophic errors, and $\Theta$ is the Heaviside step function.
The added term on the right models the fact that, of
the $\bar{n}^i$ galaxies assigned to that photometric-redshift bin, $x\bar{n}^i$ of them
have spectroscopic redshifts which are randomized across the full
range of the survey. For our analysis, we choose the range of these randomized
redshifts to be $z\in [z_{\rm  min},z_{\rm  max}]=[0.01, 2.5]$.
In practice, we significantly smooth the edges of the step
function to avoid numerical artifacts in our $\Cl$ calculations.

For this study, we use two different true
(simulation) catastrophic
photo-$z$ fractions: $x=0.01$ and 0.1; these values roughly bracket the
  currently achieved levels of catastrophic outliers in current surveys
  (e.g., CFHTLens \cite{Hildebrandt:2011hb}).
Figure~\ref{fig:catztest_rho_exp} shows the fractional change in
$\langle\rho\rangle$ and $\langle s\rangle$ when the ISW estimator is
constructed assuming various values of $x$, with true $x=0.01$ and $x=0.1$
shown in blue and brown lines, respectively.

Our results show us two things.
First,
though
mismodeling $x$ results in more significant changes than what was seen for the
survey depth and redshift-dependent bias, the shifts are still relatively
small; in the worst-case scenarios, $\langle\rho\rangle$ shifts by less than
10\% and $\langle s\rangle$ shifts by about 20\%. Second,
the constant-bias-fitting step of our pipeline does not provide protection against
mismodeled catastrophic photo-$z$ error rates. This is because the $dn/dz$
modification in Eq.~(\ref{eq:dndz_withcatz}) alters $\Cl$ in a
scale-dependent way, as can be seen in the plots in
Appendix~\ref{sec:zdisttest_cl_plots}.

To check whether catastrophic photo-$z$ errors are more damaging
when LSS data are binned in redshift, we ran a similar analysis for a case where
the fiducial $dn/dz$ was split into three redshift bins.  We observed fractional changes
in the quality statistics  similar to  those seen for the
one-bin case, so we conclude that our results are roughly independent of the binning strategy.

In summary, we find that properly modeling a survey's catastrophic photo-$z$ error
fraction is more important for preserving ISW
reconstruction quality than either its depth or redshift-dependent
bias but that, overall, reconstruction is relatively robust against these kinds of errors.

\subsection{Photometric calibration errors}\label{sec:caliberror}

Photometric calibration errors are a very general class of
systematics that cause the magnitude limit of a survey to vary
across the sky. This introduces direction-dependent number density variations which
do not correspond to fluctuations in physical matter density, thus biasing the observed
galaxy power spectrum.  Examples of photometric calibration errors
include atmospheric blurring, unaccounted-for Galactic dust, and imperfect
star-galaxy separation, among other things. A number of recent
LSS observations have found a significant excess of power at large
scales~\citep{Pullen:2012rd,  Ho:2012vy, Ho:2013lda, Agarwal:2013qta,
 Giannantonio:2013uqa,   Agarwal:2013ajb}, suggesting the presence of this
kind of  error.

We adopt a parametrization of calibration errors from
\citet{Huterer:2012zs}, who presented a systematic study of
the effects of calibration errors and requirements on their control for cosmological
parameter estimates. See also Refs.~\citep{Leistedt:2013gfa, Leistedt:2014wia, Shafer:2014wba}
for other approaches.
We model photometric calibration errors in terms of a calibration error field
$c(\nhat)$ which modifies the observed number density $N^{\rm obs}$ via
\begin{equation}\label{eq:calerror_def}
N^{\rm obs}(\nhat) = \left(1+c(\nhat)\right)N(\nhat).
\end{equation}

\begin{figure*}
  \includegraphics[width=.49\linewidth]{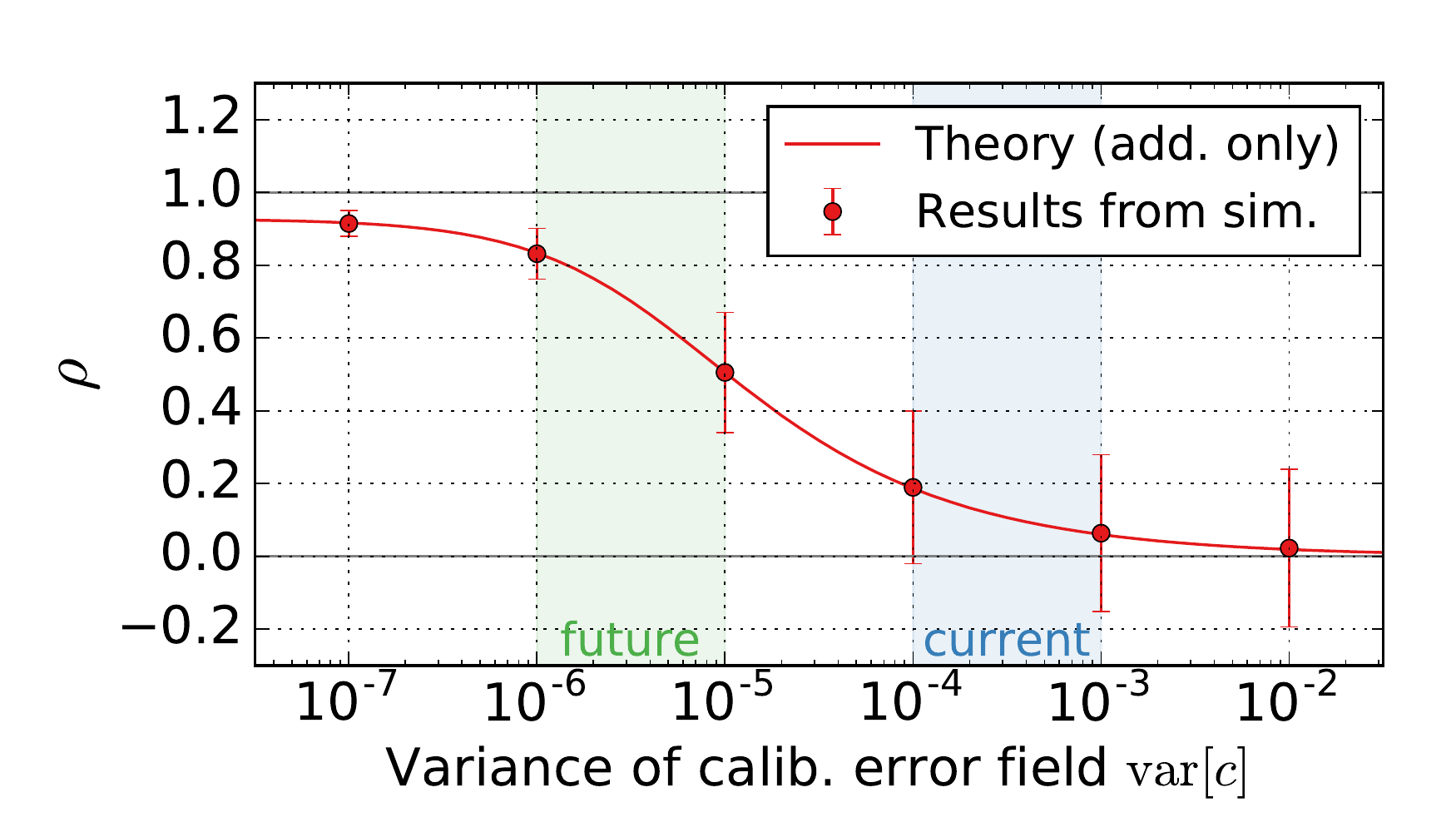}
  \includegraphics[width=.49\linewidth]{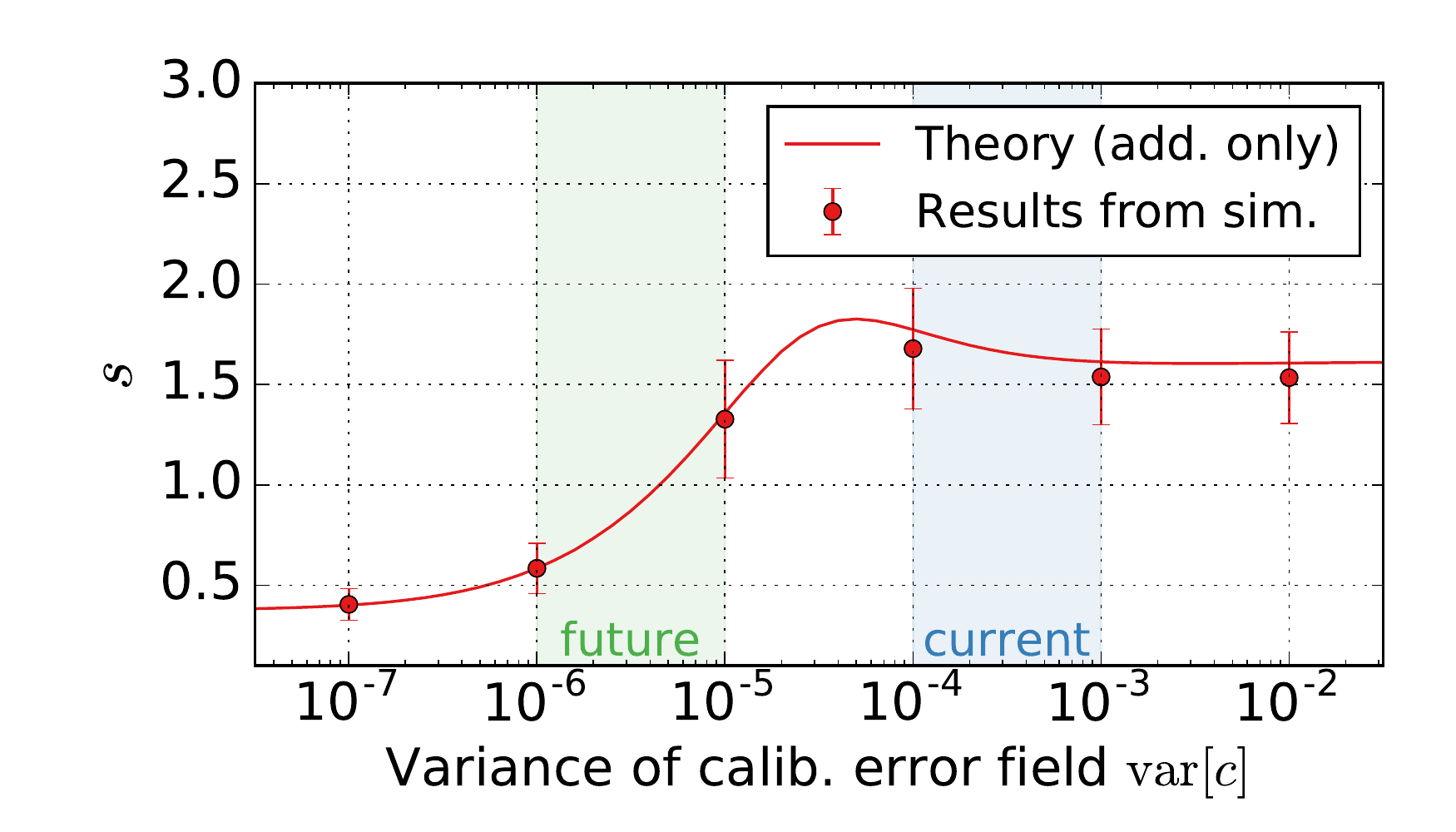}
  \caption{The effect of photometric calibration errors on reconstruction
    quality. We show results for the correlation coefficient between true and
    reconstructed ISW maps (left panel) and for the typical size of map residuals
    relative to the variance of the true ISW map (right panel). The lines show the
    expectation from theory, considering only additive contributions from calibration
    errors, while the data points show the mean and standard deviation from 10,000
    simulated map realizations. The shaded regions show the  current
    and projected levels of control over residual calibration errors discussed in
    Sec.~\ref{sec:calerror_context}.}
  \label{fig:caltest_rhoexp}
\end{figure*}

This kind of direction-dependent ``screen'' is straightforward to implement on the
level of maps but complicates the process of computing the theoretical
  expectation value for our statistics, 
$\langle\rho\rangle$ and $\langle s\rangle$.
Because multiplicative effects  introduce mixing  between
spherical components of the galaxy maps, there is a nontrivial relationship
between the power spectra for the true   galaxy distribution, the observed  galaxy distribution, and the calibration error field $c(\nhat)$. 
(See, for example, Refs.~\cite{Huterer:2012zs,Shafer:2014wba}.) To make calculations 
tractable, we use the fact that calibration error effects
will be dominated by additive contributions at large angular scales and estimate
%
\begin{equation}
  \label{eq:cl_wcalerror_addonly}
  \left[\Cl^{XY}\right]^{\rm obs}\approx
  \frac{\Cl^{XY}+C^{{\rm cal}XY}_{\ell} -
    \delta_{\ell 0}c^X_{00}c^Y_{00}}{(1+c^X_{00}/\sqrt{4\pi})(1+c^Y_{00}/\sqrt{4\pi})}.
\end{equation}
Here, $C^{{\rm cal}XY}_{\ell}$ is the cross-power between calibration error
fields affecting maps $X$ and $Y$. The $c^X_{00}\equiv \left(C^{{\rm
    cal}X}_{\ell=0}\right)^{1/2}$ terms are their monopoles, which contribute by
shifting $\bar{n}^X$. We derive this expression in
Appendix~\ref{app:calerrors}.

Note that this modification is only applied to
$\Cl^{\rm true}$. We wish to study the impact of uncorrected calibration
errors, so we will always (when analyzing simulations or calculating quality
statistic expectation values) compute $\Cl^{\rm model}$ without including
calibration error effects.

For this analysis, we adopt a functional form for 
the calibration error field power spectrum,
\begin{equation}\label{eq:clcalmodel}
  \Cl^{\rm cal} = \begin{cases} \alpha^{\rm cal} \exp{\left[-(\ell/10)^2\right]}
  &\text{ if } \ell\leq 30\\ 0 &\text{otherwise}
  \end{cases}
\end{equation}
where $\alpha^{\rm cal}$ is a normalization constant set to fix the variance of
$c(\nhat)$ to a desired value. The variance is given by
\begin{equation} {\rm var}\left[c\right] \equiv \langle c^2(\nhat)\rangle =
  (4\pi)^{-1}\sum_{\ell}(2\ell+1)\Cl^{\rm cal}.
\end{equation}
The form of Eq.~(\ref{eq:clcalmodel}) is inspired by power spectrum 
estimates for maps of dust extinction corrections and magnitude limit
variations in existing surveys. (See Figs.~5 and~6 in Ref.~\cite{Huterer:2012zs})  Using
this power spectrum, we generate independent Gaussian realizations of $c(\nhat)$
which are then combined with our simulated galaxy maps according to
Eq.~(\ref{eq:calerror_def}). These postprocessed maps are used as
input for ISW reconstruction.

\subsubsection{Context: Current and future levels of calibration error}\label{sec:calerror_context}

To put our results in context, it is useful to identify what values of
variance in the calibration field ${\rm var} [c]$ are expected from
current and future surveys. Here we emphasize that we are talking about {\it
  residual} calibration errors---that is, calibration errors which are not
properly corrected for and thus can cause biases in cosmological
inferences.

Above, we defined these errors in terms of variations in the number
of observed galaxies. To relate this to variations in a survey's limiting
magnitude, we must multiply the magnitude variations by a factor of $\ln (10)s(z)$,
where $s(z)\equiv \left .d\log_{10} N/dm \right |_{\rm m_{\rm lim}}$ is
the survey-dependent faint-end slope of the luminosity function; see Eq.~(30)
in Ref.~\cite{Huterer:2012zs}. We adopt $s(z)\simeq 0.3$ estimated from the
simulations of Ref.~\cite{Jouvel:2009mh}, assuming a median galaxy redshift $z\sim 0.75$.
This means that the conversion factor is $\ln (10) s(z)\sim 1$, and variance in
calibration is roughly equal to that in the limiting magnitude,
$c(\nhat)\equiv (\delta N / N) (\nhat)\simeq (\delta m)_{\rm lim}$.

With these assumptions, the smallest currently achievable variance of the
calibration error $c(\nhat)$ is of order ${\rm var} [c]\sim 10^{-3}$
(e.g., Fig.~14 in Ref.~\cite{Leistedt:2013gfa}).  For example, residual limiting
magnitude variations in the SDSS DR8 survey are at the level of $0.03$ mag
\cite{2015arXiv150900870R}, again implying that ${\rm var} [c]\simeq
10^{-3}$.  Note that, while the impressive SDSS ``uber-calibration'' to 1\%
\cite{Padmanabhan:2007zd} would imply an order of magnitude smaller variance,
this might be difficult to achieve in practice because there are sources of
calibration error that come from the analysis of the survey and are not
addressed in the original survey calibration.  We show the current levels of
residual calibration errors value as a blue vertical band in
Fig.~\ref{fig:caltest_rhoexp}, spanning a range between the optimistic level
associated with the SDSS uber calibration to the more conservative ${\rm  var}[c]=10^{-3}$.

In the same figure, we also show the {\it future} control of calibration
errors required to ensure that they do not contribute appreciably to
cosmological parameter errors---e.g., those in dark energy and primordial
non-Gaussianity. This range, forecasted assuming final DES data and adopted
from Ref.~\cite{Huterer:2012zs}, is shown as a green band spanning ${\rm
  var}[c]\sim 10^{-6}$--$10^{-5}$.  The lower bound is set by the
requirement that the bias to cosmological parameter estimates be smaller than
their projected errors, while $10^{-5}$ is chosen as an intermediate value
between that and ${\rm var}[c]=10^{-4}$, which introduces unacceptable levels
of bias. (See Fig.~4 of Ref.~\cite{Huterer:2012zs}.) These should be viewed as only
rough projections, as the precise requirements depend
on the faint-end slope $s(z)$ of the source luminosity function, the
cosmological parameters in question, and the shape of the calibration field's
power spectrum $\Cl^{\rm cal}$.
  
\subsubsection{Results for ISW reconstruction}\label{sec:calerror_results}

We find that even small levels of calibration error can have a significant impact on
ISW reconstruction quality. Figure~\ref{fig:caltest_rhoexp} shows how the correlation
between true and reconstructed maps, $\rho$, and the reconstructed map
residuals, $s$, respond to different levels of calibration error.

Reconstruction quality starts to degrade when ${\rm var}[c]\sim 10^{-6}$,
which roughly corresponds to the same 0.1\% magnitude
  calibration  required to achieve cosmic-variance-limited ISW
  detection~\cite{Afshordi:2004kz}.
At this
level, we see $\rho$ begin to move away from its best-case (no calibration error)
value and the $s$ plot shows that residuals are comparable in
amplitude to fluctuations due the true ISW signal. 

Once the calibration error power starts to dominate over the galaxy autopower,
occurring around  ${\rm var}[c]\sim 10^{-4}$, the reconstruction contains little
information about the true ISW signal. Here, the scatter in $\rho$ overlaps with
zero and we see that the reconstructed map residuals approach a constant value.
See Appendix~\ref{app:calerrors_slim} for an explanation of why we expect this
to occur. 

Comparing these numbers to the shaded bands, we see that, with current levels
of calibration error control, we have little hope of accurately reconstructing
the ISW signal with galaxy survey data alone. Encouragingly, though, the levels
of control required to obtained unbiased cosmological parameter estimates from
next-generation surveys~\cite{Huterer:2012zs} are precisely the levels needed
for accurate ISW reconstruction. 

We note that the additive-error-only theory calculations show good agreement
with our results from simulations, and so can 
be useful as a computationally efficient indicator of when calibration errors
become important.  In light of this,  we also
computed $\langle \rho \rangle$ and $\langle s \rangle$ using a
power law spectrum, $\Cl^{\rm cal}\propto \ell^{-2}$, in order to check how
sensitive our results are to the shape of the calibration error field's power
spectrum. This more sharply peaked spectrum caused reconstruction
quality to start degrading at a slightly smaller ${\rm var}[c]$ compared to the
Gaussian model, but otherwise showed similar results. This can likely be explained by
the fact that  the power law $\Cl^{\rm cal}$ reaches higher values at low $\ell$
for a given
field variance, which means it can start dominating over true
galaxy power at those multipoles earlier. 


\subsubsection{Mitigation by raising $\ell_{\rm min}$}
Because calibration error fields tend to have the most power on large scales,
we looked at whether raising $\ell_{\rm min}$ can mitigate their impact. Our
results, shown in Fig.~\ref{fig:caltest_lminplot}, show that raising
$\ell_{\rm min}$ from 2 to 3 or 5 causes the error bars denoting the scatter in
$\rho$ to cross zero at a higher value of ${\rm var}[c]$. However, this
effect is small, and we conclude that raising $\ell_{\rm min}$ provides only
limited protection against calibration errors.

\begin{figure}
 \includegraphics[width=1.0\linewidth]{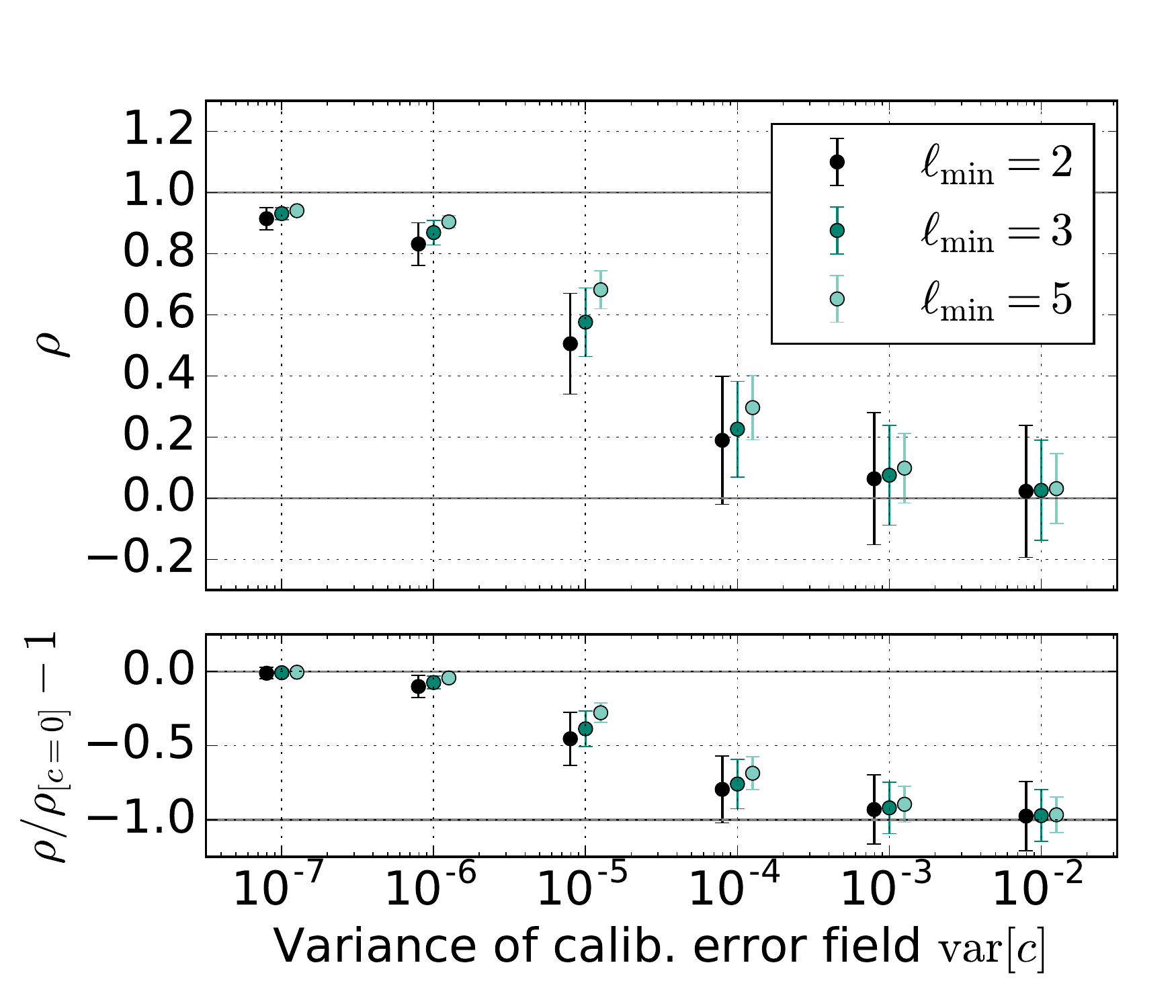} 
  \caption{Exploration of whether raising $\ell_{\rm min}$ can mitigate the
    impact of photometric calibration errors on ISW signal reconstruction. The
    top panel shows the mean and standard deviation of $\rho$, the correlation
    between the true and reconstructed ISW maps, measured from 10,000
    simulations. The bottom panel shows the fractional change in $\rho$
    relative to the case with no calibration errors. Points for different
    values of $\ell_{\rm min}$ are staggered so that the errors bars are
    legible; each cluster of three points shares the same value of ${\rm
      var}\left[c\right]$.}
  \label{fig:caltest_lminplot}
\end{figure}

\section{Implications for cosmic alignments}\label{sec:anomalies}

Over the past 15 years, as the full-sky CMB maps provided by the WMAP and Planck
experiments became available, increasing evidence has been found for anomalies
at large angular scales. In particular, angular correlations at scales above
60 deg on the sky seem to be missing, while the the quadrupole and
octupole moment of the CMB anisotropy are aligned both mutually and with the
geometry and the direction of motion of Solar System. The origin for the
anomalies is not well understood at this time; they could be caused by
astrophysical systematic errors or foregrounds or cosmological causes (like
departures from simple inflationary scenarios), or they could be a
statistical fluctuation, albeit a very unlikely one. The anomalies have most
recently been reviewed in Ref.~\cite{Schwarz:2015cma}.

Some authors \cite{Francis:2009pt,Rassat:2013caa} have commented
on the fact that current efforts to ``peel off'' the ISW contribution
from the CMB maps indicate
that the significance of some CMB anomalies is
``significantly reduced'' once the ISW contribution is subtracted.
If true, this statement
implies that the observed anomalies are either due to features in the ISW map
or caused by an accidental alignment of the early- and late-time CMB
anisotropy \cite{Copi:2013jna}.  In any case, statements on how the
primordial and late CMB combine to produce the anomalies clearly depend on
the fidelity of the reconstructed ISW contribution to the CMB, which is the
subject of our work.

Our goal here is not to carry out a full investigation of the ISW
map reconstruction's effect on the anomalies' significance. Instead,
we would like to
simply build intuition on how much imperfect reconstruction affects 
inferences about the anomalies.

To that end, we pose the following question: if we {\it assume} for the
moment that an ISW map reconstructed using available LSS data happens to show
a significant quadrupole-octupole alignment, what is the likelihood  that the
true ISW map is actually aligned? Note that we in no way imply that the ISW-only
alignment scenario is
 a favored model for the observed CMB anomalies. We simply want to
study how robust certain properties of the ISW map, particularly the phase
structure of the anisotropies in the map, are to the
reconstruction process.

To study the alignments, we adopt the (normalized) angular momentum dispersion
maximized over directions on the sky, defined as \citep{deOliveira-Costa:2003utu,Copi:2005ff}
\begin{equation}
  (\Delta L)^2_{\rm 2+3, true}\equiv
  \max_{\nhat}
  \left (\frac
  {\sum_{m=-\ell}^\ell m^2\, |\alm (\nhat)|^2}
  {\ell^2\sum_{m=-\ell}^\ell\, |\alm (\nhat)|^2}
  \right )
  \label{eq:Lmax}
\end{equation}
where  $\alm (\nhat)$ are expansion coefficients of the map in a coordinate system
where the $z$-axis is in the $\nhat$ direction. Hence, the maximization is
performed over all directions $\nhat$; note that only the numerator of the
expression in angular parentheses depends on the direction, and see Sec.~5.6 of
Ref.~\cite{Copi:2005ff} for the algorithm to efficiently compute the
maximization. Intuitively, high values of the angular momentum indicate
significant planarity of the $\ell=2$ and $\ell=3$ modes as well as their
mutual alignment.  

We set up the following pipeline:
 \begin{itemize}
  \item Start with $10,000$ random realizations of the true ISW map and the
    corresponding LSS maps (so that each LSS map contains gravitational
    potential field that produces the corresponding ISW map).
  \item For each true ISW map, measure the angular momentum dispersion
    $(\Delta L)^2_{\rm 2+3, true}$ defined in Eq.~(\ref{eq:Lmax}).
  \item Reconstruct each map assuming a fiducial LSS survey and repeat
    the calculation to get a set of $(\Delta L)^2_{\rm 2+3, rec}$.
  \item Make a scatter plot of $(\Delta L)^2_{\rm 2+3, rec}$ vs $(\Delta
    L)^2_{\rm 2+3, true}$, which will show  how much and in
          which direction reconstruction biases the alignment information.
  \end{itemize}

\begin{figure}
 \includegraphics[width=1.1\linewidth]{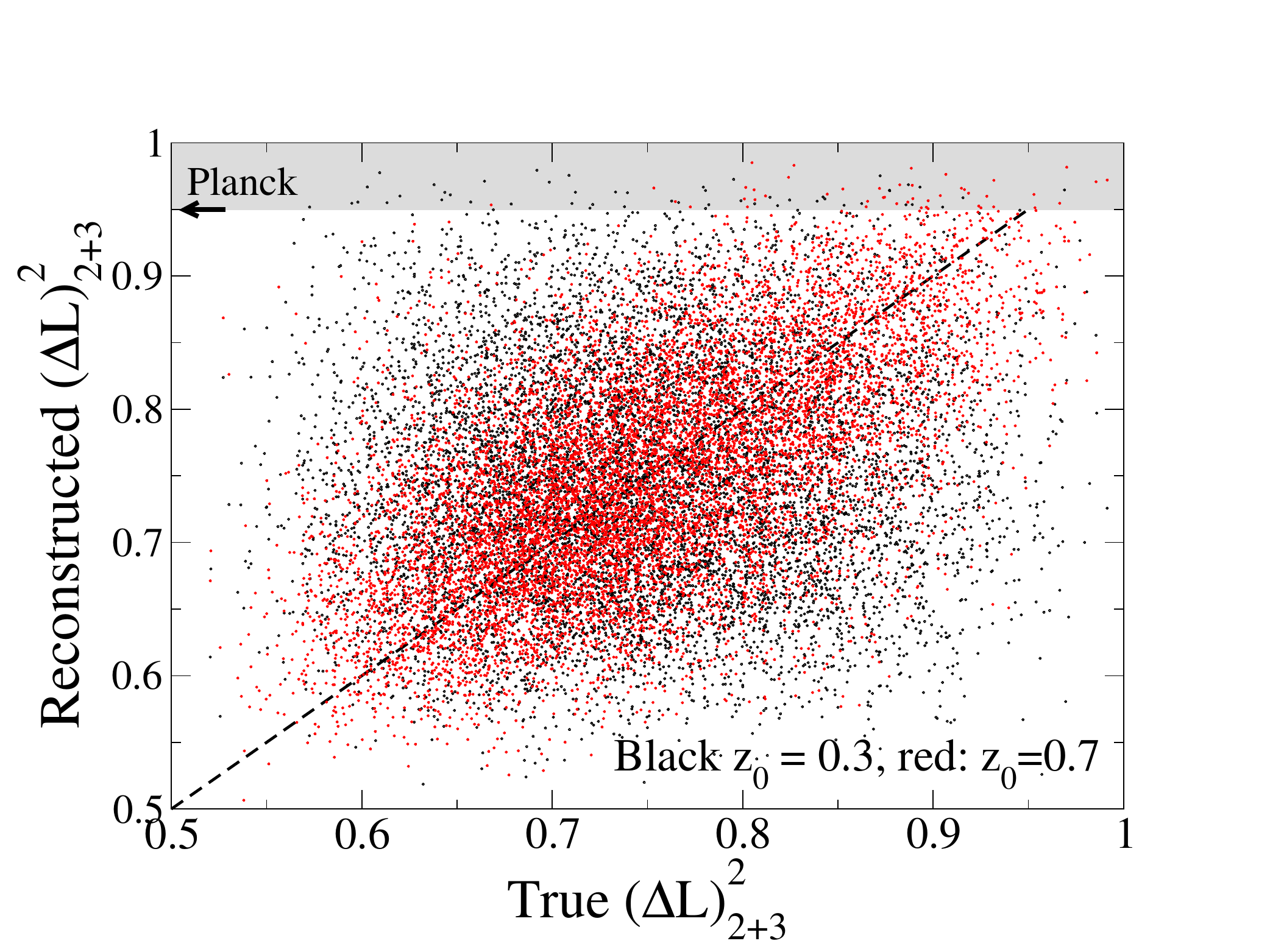} 
 \caption{The relationship between the true ($x$-axis) and reconstructed ($y$-axis)
   angular momentum dispersion $(\Delta L)^2_{\rm 2+3}$, defined in the text, for the
   combined quadrupole and octupole in 10,000 randomly generated ISW maps.
   Results are shown for two alternate survey depths: our fiducial LSS survey with
   $z_0=0.7$ (red points) and  $z_0=0.3$ (black points), which have correlation
   coefficients 0.58  and  0.11 respectively.
   The  gray region denotes $(\Delta L)^2_{\rm 2+3}$ as high or higher
   than  measured in WMAP and Planck CMB maps, while the diagonal line is
   where the true and reconstructed values match. See the text for details.}
\label{fig:angmom}
\end{figure}

The results are summarized in Fig.~\ref{fig:angmom}. There we show how the
inferred angular momentum dispersion of the combined quadrupole and octupole
is affected by reconstruction for 10,000 randomly generated ISW maps. The
$x$-axis shows the value for the is the true ISW map, while the $y$-axis shows
values reconstructed from our fiducial LSS survey at two alternate depths,
$z_0=0.7$ (red points) and 0.3 (black points). We find that the true and
reconstructed angular momentum dispersions are not very correlated, having a
correlation coefficient of only 0.58 for $z_0=0.7$ and 0.11 for $z_0=0.3$.

We
also denote the value for the angular momentum dispersion of the WMAP/Planck
full map, which includes both primordial and late-time ISW contributions, at
$(\Delta L)^2_{\rm 2+3}\simeq 0.95$.  (The precise value varies slightly
depending on the map.~\cite{Copi:2005ff,Schwarz:2015cma}) Of the $z_0=0.7$
(0.3) reconstructed maps which have $(\Delta L)^2_{\rm 2+3}$ as high as or
higher than the WMAP and Planck CMB maps (points falling in the shaded gray
region), only 10\% (2\%) have corresponding true maps which satisfy the same
high angular momentum dispersion criterion.

Investigating the implications of the ISW reconstruction on the
inferences about the alignments of primordial-only and ISW-only maps in depth
is beyond
the scope of this paper. Nevertheless, our simple test indicates that at least
the quadrupole-octupole alignment {\it in the ISW-only maps} is not very robust under ISW reconstruction
using realistic LSS maps, even without taking into account calibration
and other systematic errors.

\section{Conclusion}\label{sec:concl}

In this work we use simulated ISW and LSS maps to study the accuracy of ISW
signal reconstructions performed using LSS data as input. In particular, we
study how systematics associated with galaxy surveys affect the ISW map
reconstruction. We measure reconstruction accuracy using two quality
statistics: $\rho$, the correlation coefficient between the true and
reconstructed ISW maps, and $s$, the rms error in the reconstructed map
relative to the rms of true ISW map features.

In the absence of systematics, we find that increasing survey depth improves
these statistics (brings $\rho$ closer to 1 and lowers $s$), though the shifts
in their average values are small compared to their scatter. Similarly,
splitting the survey data into redshift bins leads to moderate
improvement. The reconstruction quality improvement due to increasing survey
depth by $\Delta z=0.1$ is comparable to that gained by splitting into three
redshift bins: both lead to improvement $\Delta\bar{\rho}\sim 0.02$, or
$\Delta\bar{\rho}/\bar{\rho}\sim 2\%$. We also find that reconstruction can be
slightly improved if we are willing to neglect the reconstruction of very
low-$\ell$ multipoles; increasing our fiducial $\ell_{\rm min}=2$ to 5 results
in $\Delta\bar{\rho}\sim 0.01$ and a reduction in the scatter of $\rho$ by
about a factor of 2. Last, we find that galaxy shot noise has a negligible
impact as long as $\bar{n}\gtrsim 1 \,{\rm arcmin}^{-2}\approx 10^7 \,{\rm sr}^{-1}$. These results provided a baseline comparison for our
studies of systematics.

The first class of systematics we study are those associated with mismodeling
the line-of-sight distribution of LSS sources. By examining what happens to
reconstruction quality when different galaxy window functions are used for the
ISW-estimator input $\Cl^{\rm model}$ than for the simulation-generating
$\Cl^{\rm true}$, we find that ISW signal reconstruction is robust against
these kinds of errors. We study the mismodeling of survey depth and
redshift-dependent bias and find that fractional shifts in
$\langle\rho\rangle$ are less than $\mathcal{O}(10^{-4})$ for all but the most
extreme cases. Inaccurately estimating the fraction of catastrophic photo-$z$
errors results in a larger shift, which depends on the true fraction, but at
worst this degrades $\langle\rho\rangle$ by about a percent.
Reconstruction quality is likely to be similarly insensitive to 
other direction-independent modeling uncertainties; for example, the choice of
cosmological parameter values and maybe models of modified gravity.

The fact that we fit data for a constant galaxy bias is the key to this
robustness. This is because the modeling errors discussed above change the
galaxy spectrum by a mostly scale-independent amplitude which is degenerate
with a shift in constant bias $\bar{b}$.  Thus, the more a given systematic changes the
shape (rather than amplitude) of galaxy $\Cl$, the more of an impact it will
have on ISW signal reconstruction.

We find that photometric calibration errors  are
by far the most important systematic to control if one wants to construct a
map of the ISW signal from LSS data. For the reconstructed ISW map to contain
accurate information about the true ISW signal, calibration-based variations
in number density must be controlled so that the calibration error field $c$,
defined via $N^{\rm obs}(\nhat)=(1+c(\nhat))N(\nhat)$, has a variance less than
$10^{-4}$. Even at that level, which is optimistic for current surveys, the
reconstruction quality is significantly degraded compared to the case with no
systematics. For the model we studied, in order to keep that degradation
smaller than $\mathcal{O}(10\%)$, calibration errors must be
controlled so that ${\rm var}[c]\lesssim 10^{-6}$. This is a similar level to
what is required to avoid biasing cosmological parameter estimates made with
future survey data. Prospects for mitigation of these effects by neglecting
low~$\ell$ multipoles are  limited.

We also briefly explore the viability of using reconstructed ISW maps to
comment on the significance and origins of observed large-angle CMB anomalies.
We do this by comparing the level of
alignment, parametrized in terms of angular momentum dispersion, observed for
the $\ell=2,3$ modes of true and reconstructed ISW maps. We find that, even in
the absence of systematics, the amount of alignment was only weakly correlated
between these maps. For example, the values of true and reconstructed  angular
momentum dispersion had a correlation coefficient of only 0.58 for our fiducial
survey. Therefore,  recovering precise  alignments of structures in the ISW map,
using only LSS data as input, seems like a very challenging prospect.

These results have implications for current and future attempts to reconstruct
the ISW signal. Most significantly, they tell us that understanding the level
and properties of residual calibration errors in LSS maps is vital to
assessing the accuracy of reconstructions made using those maps as input. Given the current levels of calibration error control, at face value our
results would seem to imply that reconstruction using existing data is
hopeless. Thus, a productive avenue for future work would be
 to modify the ISW reconstruction pipeline to make
it more robust against calibration errors, by including them in the ISW
estimator's noise modeling or by some other method. Since the presence of uncorrected calibration errors will cause one to underestimate galaxy-galaxy noise, it would also be worth turning a critical eye toward how calibration uncertainties affect the evaluation of ISW detections' signal to noise.

We note that using multiple cross-correlated LSS data sets---which map the same potential fluctuations but are presumably subject to
different systematics---will mitigate the impact of calibration
errors, as will combining LSS maps with CMB temperature and
  polarization data. The results of the binning test in Sec.~\ref{sec:bintest}
 provide provisionary evidence for this, though for that
 study it is not possible to disentangle the effects of noise mitigation from
 those of adding tomographic information. An interesting extension to this
work would thus be to explore in more detail
whether and to what extent using multiple LSS maps protects ISW
reconstruction against calibration errors. Studying the combination of multiple surveys introduces a number of new questions: one might study, for example, how the strength of correlation between galaxy maps influences the improvement in reconstruction due to their combination, or what happens when calibration errors for multiple maps are correlated. In order to give these questions their due attention, and for the same of conciseness, we defer this study to a followup paper.


\acknowledgements The authors have been supported by DOE under Contract No. DE-FG02-95ER40899. D. H.
has also been supported by NSF under Contract No. AST-0807564, NASA, and DFG Cluster of
Excellence “Origin and Structure of the Universe”
(http://www.universe-cluster.de/).  We thank Max Planck Institute for
Astrophysics for hospitality.

\appendix

\section{Cross-check with~\citet{Manzotti:2014kta}}\label{app:MDcrosscheck}

\begin{figure*}
 \includegraphics[width=.5\linewidth]{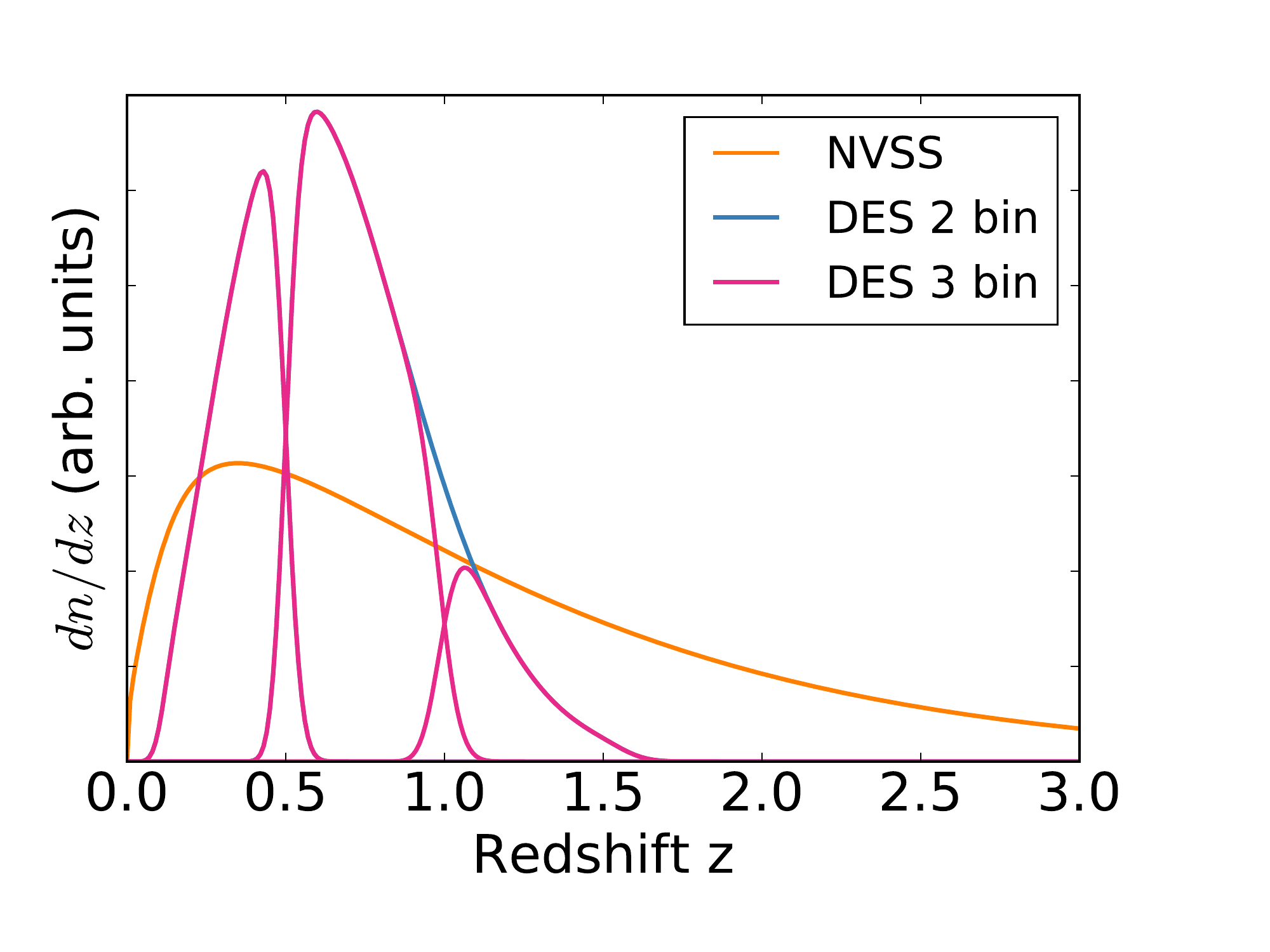}\includegraphics[width=0.5\linewidth]{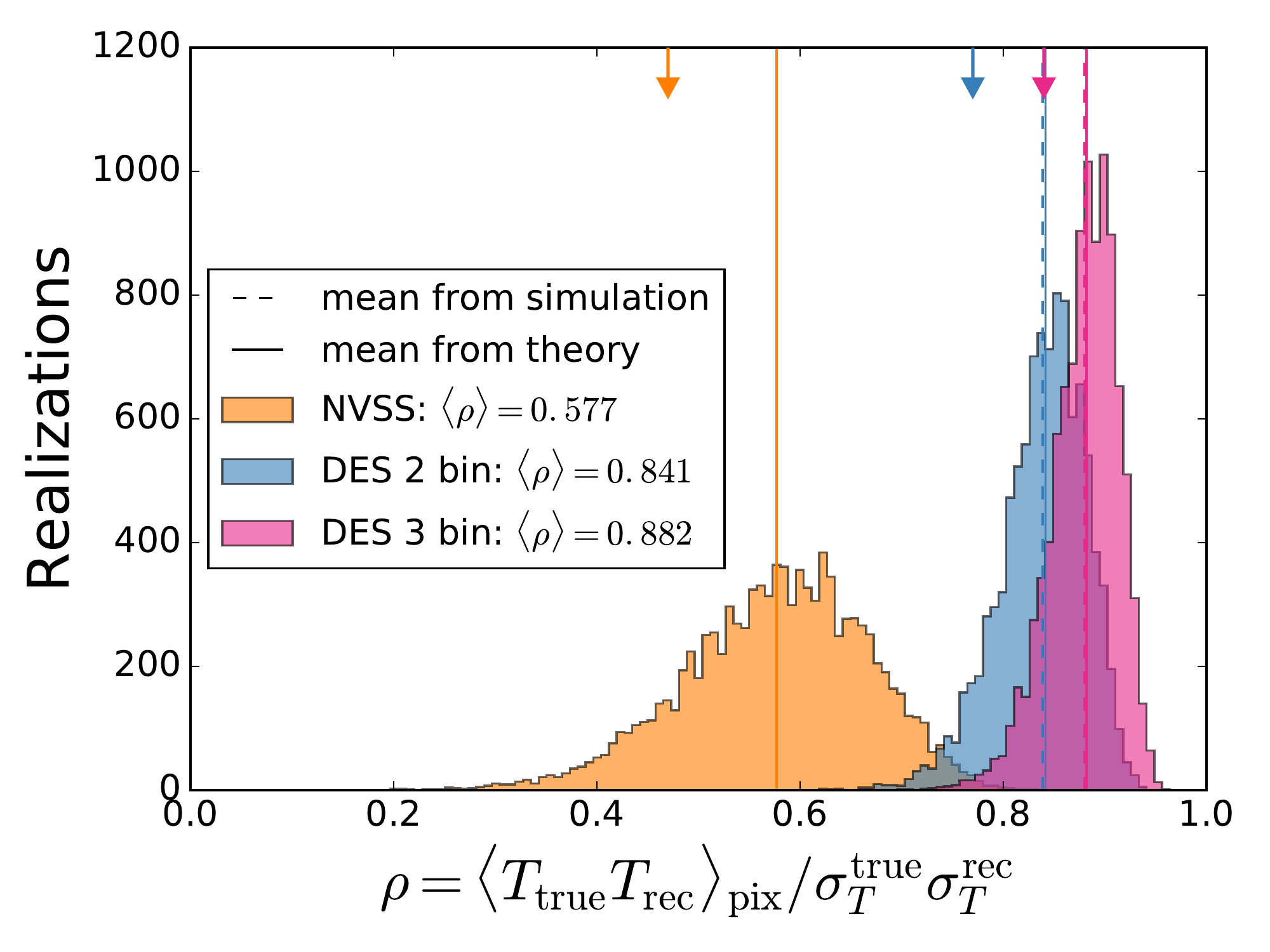}
 \caption{Left panel: Redshift distributions of surveys, chosen to match the LSS surveys
   studied in Ref.~\cite{Manzotti:2014kta}. Right panel: Histogram of $\rho$ found for 10,000
   simulations of   surveys with redshift distributions shown in the left panel. Values
   of $\bar{\rho}$ from Ref.~\cite{Manzotti:2014kta} are shown
   by the arrows along the top of the plot. The observed discrepancies are
   likely due to different amounts of simulated galaxy shot noise.}
  \label{fig:MDcheck_dndz_rhohist}
\end{figure*}

Here we perform a crosscheck of our reconstruction procedure
against~\citet{Manzotti:2014kta} (MD). In their paper, MD
perform simulations for an NVSS-like survey and a DES-like survey in two- and
three-binned configurations. We attempt to simulate ISW reconstruction for
similar surveys.

For the NVSS-like survey, we use the analytic $dn/dz$ distribution given by
MD, integrating between $0.01\leq z\leq 6$ when computing its $\Cl$.
The redshift distributions used for these simulations
are shown in the left panel of Fig.~\ref{fig:MDcheck_dndz_rhohist}. For the
DES-like survey, we adjusted the parameters in our fiducial $dn/dz$ model by
eye so that the three-binned case is similar to that shown in MD's relevant
figure.  For the three-binned case, we place bin edges at
$z\in[0.1,0.5,1.0,1.6]$. Because MD do not describe how the two-binned case is
divided, we somewhat arbitrarily place the bin edges at
$z\in[0.1,0.5,1.6]$.  Like MD,
we include multipoles $3\leq \ell\leq 80$ in our analysis. We leave $\bar{n}$
at our fiducial value of $10^9$ for all of these surveys. This value was
selected based on an assumption that shot noise contributions would be
negligible, but we note below that this is likely not the case.

The right panel of Fig.~\ref{fig:MDcheck_dndz_rhohist} shows a histogram of
the $\rho$ values for 10,000 map realizations in our study, with the values
from MD shown with arrows. We find that our $\bar{\rho}$ values are
systematically higher than, but not wildly incompatible with those in MD. It is
hard to specifically identify a cause for this without more information, but
the discrepancy is most likely due to differences in the amount of Poisson
noise we add to our galaxy maps. We note, for example, that we can get our
$\langle\rho\rangle$ for the NVSS-like survey to roughly match the MD value if
we reduce our simulation's $\bar{n}$ to $\sim 5\times 10^5$. If we set
$\bar{n}$ to the value reported for NVSS by MD, $\bar{n}=5\times 10^4 \,{\rm
  sr}^{-1}\approx 16 \,{\rm deg}^{-2}$, we get a lower value of
$\langle\rho\rangle=0.22$.


The shift between the two- and three-bin DES surveys in our simulations is larger than
the $\Delta\rho\sim0.03$ seen in the binning study of Sec.~\ref{sec:bintest}. This
supports our hypothesis that $\bar{\rho}$ shifts more easily at lower $\rho$ values.
The fact that our observed shift is still only about half the size of that
by MD is probably also due to the fact that we are finding larger
$\bar{\rho}$ values than they do.


\section{$\Cl$ plots for Sec.~\ref{sec:zdisttest}}\label{sec:zdisttest_cl_plots}
Figure~\ref{fig:zdisttest_cl} shows how
galaxy-galaxy and galaxy-ISW power spectra respond to changes in the parameters
discussed in Sec.~\ref{sec:zdisttest}. We study the effect of survey depth by shifting
the parameter $z_0$ in Eq.~(\ref{eq:fiddndz}), redshift dependence of bias by
changing $b_2$ in Eq.~(\ref{eq:bzmodel}), and the fraction of galaxies $x$ subject
to catastrophic photometric-redshift errors via Eq.~(\ref{eq:dndz_withcatz}).

We see that changing $z_0$ and $b_2$ shifts $\Cl$ by a mostly scale-independent factor.
As noted in Sec.~\ref{sec:zdisttest}, this is why  systematics related
to mismodeling depth and bias redshift dependence have only a small effect on
ISW reconstruction quality. It is also why fitting for  scale-independent bias
$\bar{b}$   via
\begin{equation}
C_{\ell}^{\rm gal (obs)}=\bar{b}^2\, \Cl^{\rm gal(model)},
\end{equation}
as is discussed in Sec.~\ref{sec:biasfitting}, protects against these systematics. 

In contrast, changing the catastrophic
photo-$z$ fraction $x$ by more than about 0.01 significantly changes the
low-$\ell$ shape of $\Cl$. This explains why mismodeling $x$ has a relatively
larger (though still small) impact on ISW reconstruction quality and why
constant bias fitting does not mitigate this effect as much.

\begin{figure*}
\includegraphics[width=.33\linewidth]{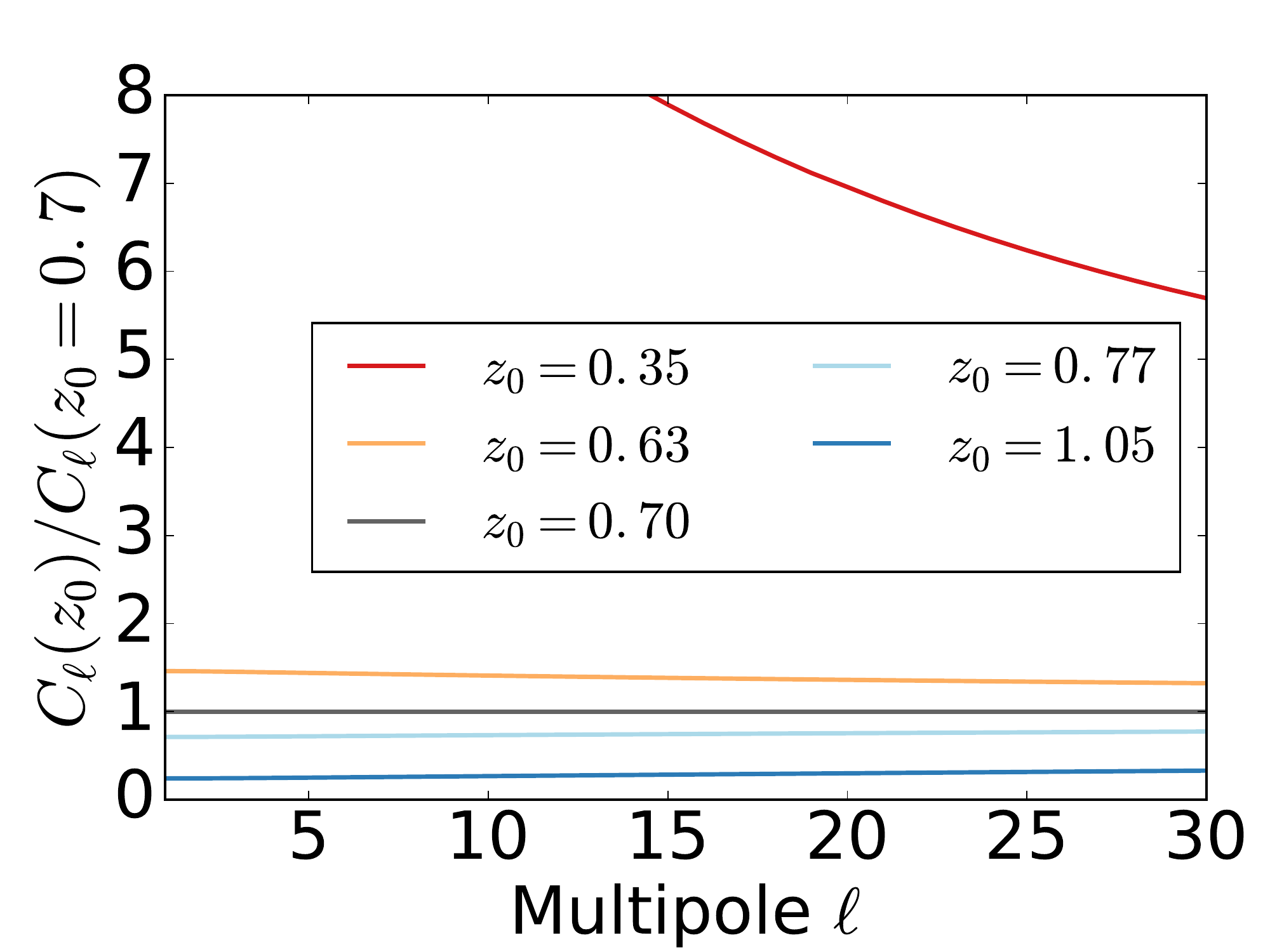}\includegraphics[width=.33\linewidth]{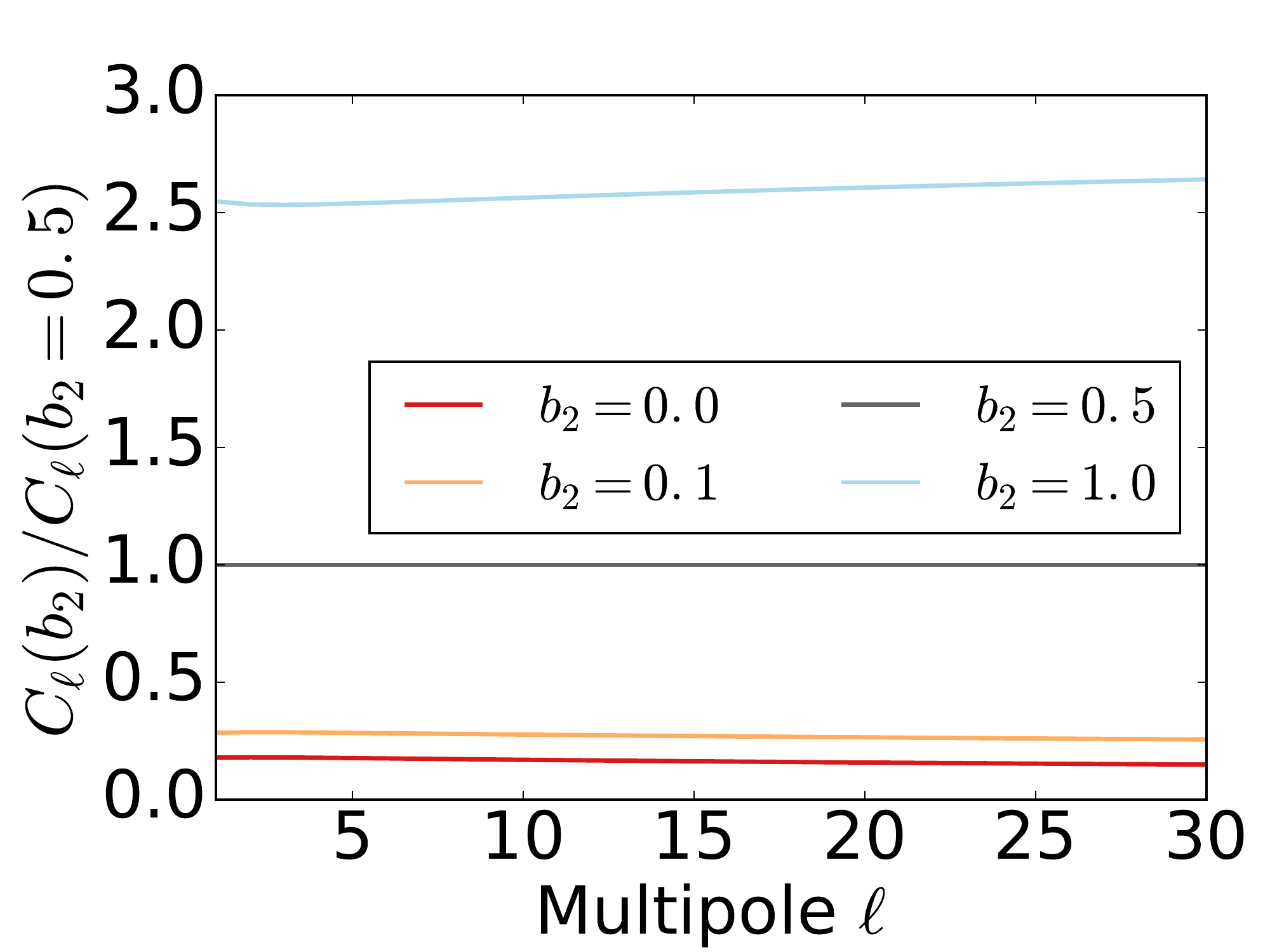}\includegraphics[width=.33\linewidth]{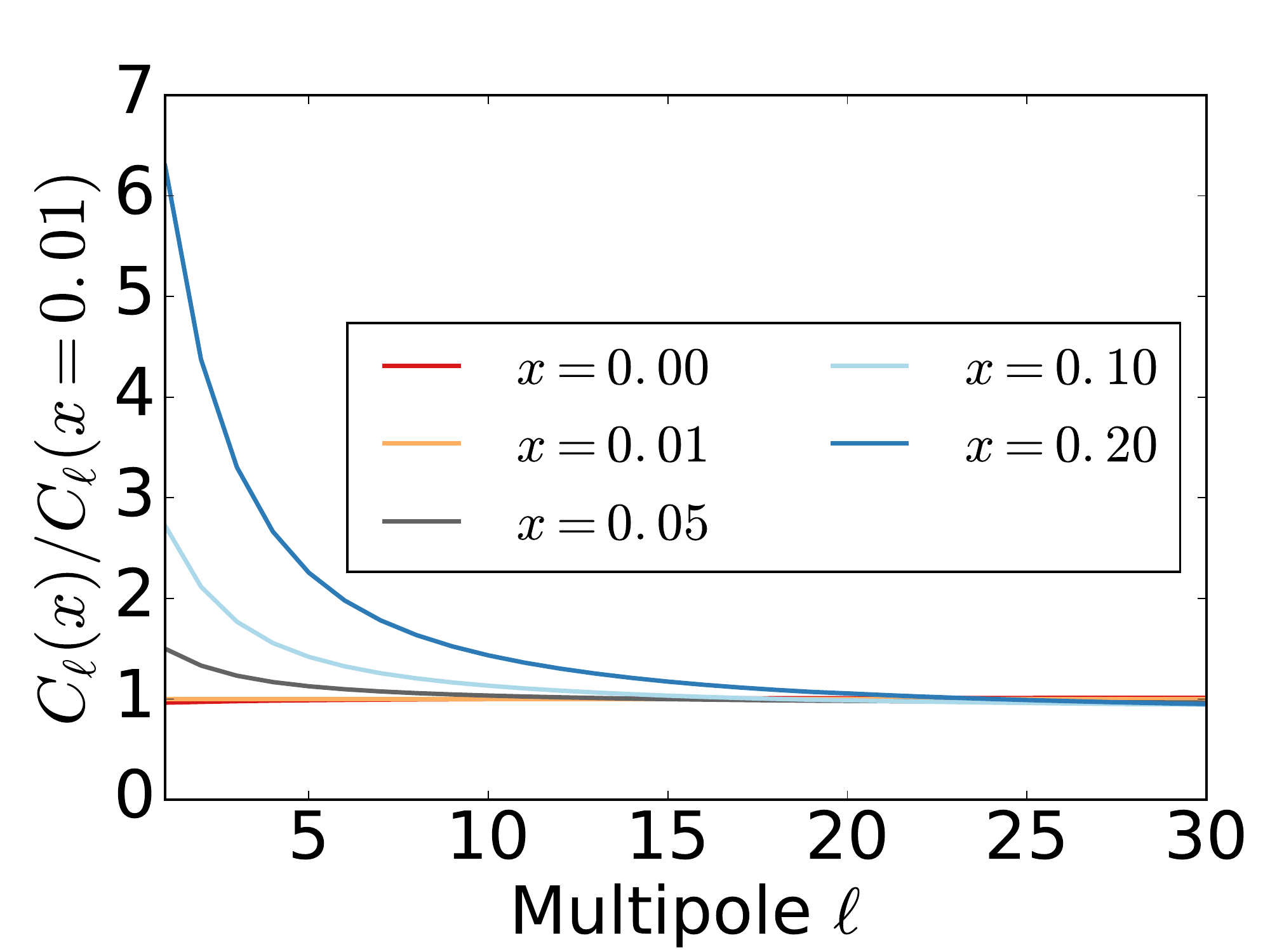} 
\caption{The change in the galaxy angular power spectrum
  $\Cl\equiv \Cl^{\rm gal-gal}$  in response to (left to right) changes in
    survey depth, characterized via $z_0$ in Eq.~(\ref{eq:fiddndz}); the
    redshift dependence of bias, modeled by varying the $b_2$ parameter in
    $b(z)=1+b_2(1+z)^2$; and the fraction $x$ of galaxies subject to
    catastrophic photo-$z$ errors. These plots show the ratio of galaxy autopower
    relative to that of a reference survey.}
  \label{fig:zdisttest_cl}
\end{figure*}

\section{Calibration error formalism}\label{app:calerrors}

In Sec.~\ref{sec:caliberror} we study the impact of photometric calibration
errors on ISW signal reconstruction. We model them using a direction-dependent
calibration error field $c(\nhat)$ via
\be
N^{\rm obs}(\nhat) = (1+c(\nhat))N(\nhat),
\ee
where $\nhat$ is the direction on the sky,  $N^{\rm obs}$ is the observed number of
galaxies, and $N$ is the true number of galaxies. Here, we present the calculations
necessary to describe how this  modifies the galaxy $\Cl$ and which we used
above to predict how calibration errors will impact our reconstructions
quality statistics. Our notation follows that by~\citet{Huterer:2012zs}.

We will define fluctuations in the true
and observed number density as $\delta$ and $\delta^{\rm obs}$, respectively, and
write them in terms of spherical components,
\begin{align}
  \delta(\nhat) &= \frac{N(\nhat)}{\bar{n}} -1 \equiv \sum_{\ell m}\glm \Ylm(\nhat)\\
  \delta^{\rm obs}(\nhat) &= \frac{N^{\rm obs}(\nhat)}{\bar{n}^{\rm obs}} -1 \equiv \sum_{\ell m}\tlm \Ylm.
\end{align}
Additionally, we will define a parameter $\epsilon$ to relate the true and
observed average number densities,
\be
\bar{n}^{\rm obs} = \bar{n}(1+\epsilon),
\ee
and use $c_{\ell m}$ to denote the spherical components of the calibration
error field $c(\nhat)$. Each galaxy map can have its own calibration error
field, and so we will use superscripts (e.g.,  $\glm^i$, $\clm^i$, and
$\tlm^i$) to denote components associated with LSS map $i$.

Our goal is to find a relation between the observed galaxy power $T_{\ell}^{ij}$,
the true power $\Cl^{ij}$, and the properties of the calibration error field
$C_{\ell}^{{\rm cal,} ij}$. To do this, we start by relating the spherical
components of the fields. We note that observed number density fluctuations
are
\be
\delta^{\rm obs}(\nhat) =  \frac{\delta+c+\delta c -\epsilon}{(1+\epsilon)},
\ee
where we suppress the $\nhat$ arguments to simplify notation. After some
algebra, we can write
\begin{align}
\tlm^i = &(1+\epsilon^i)^{-1}\left[-\sqrt{4\pi}\delta_{\ell 0}\epsilon^i+\glm^i+\clm^i\right. \\&\left.+\sum_{\ell_1\ell_2 m_1m_2}\,c^i_{\ell_2 m_2}g^i_{\ell_1 m_1}\,R^{\ell\ell_1\ell_2}_{mm_1m_2}\right].
\end{align}
In this expression, $\delta_{\ell 0}$ is a Kronecker delta, and  the multiplicative term
\be
R^{\ell\ell_1\ell_2}_{mm_1m_2}\equiv \int\,d\Omega \Ylm^*(\nhat)\,Y_{\ell_2 m_2}(\nhat)\,Y_{\ell_1 m_1}(\nhat)
\ee
is related to Wigner-3j symbols. 

We define the cross-power between two observed maps via
\be
T_{\ell}^{ij} \equiv \sum_m \frac{\langle \tlm^i t_{\ell m}^{j*}\rangle}{2\ell +1}
\ee
and that of the calibration error fields  as
\be
C_{\ell}^{{\rm cal,} ij} \equiv \sum_m \frac{\clm^i c_{\ell m}^{j*}}{2\ell +1}.
\ee
Note that these definitions do not preclude the possibility that the $\clm$
could introduce correlations between different ($\ell, m$) modes. The fact
that we only show correlations between modes with matching $\ell$ and $m$
reflects the (potentially biased) measurement that would be made even if one
assumes that they do not.

The expression for $T_{\ell}^{ij}$ in terms of $\glm$, $\clm$ is fairly involved,
though it can be simplified to some extent using Wigner-3j symbol
identities. For the purposes of this paper, we approximate it by only
including additive components---that is, neglecting all terms containing
$R^{\ell\ell_1\ell_2}_{mm_1m_2}$. Doing this, and using the fact that
\be
\langle\epsilon^i\rangle = \frac{ c_{00}^i}{\sqrt{4\pi}} = \sqrt{\frac{C^{{\rm cal,}i}_{\ell=0}}{4\pi}} ,
\ee
we write
\be
 T_{\ell}^{ij} = \frac{\Cl^{gij}+\Cl^{cij}-\delta_{\ell 0}c^i_{00}c^j_{00}}{(1+c_{00}^i/\sqrt{4\pi})(1+c_{00}^j/\sqrt{4\pi})}.
 \ee
This is the expression given in Eq.~(\ref{eq:cl_wcalerror_addonly}) and is
what is used to compute expectations values of ISW reconstruction quality
statistics in Sec.~\ref{sec:caliberror}.

\section{Large-noise limit of $s$ statistic}\label{app:calerrors_slim}

In Sec.~\ref{sec:calerror_results}, and particularly in
Fig.~\ref{fig:caltest_rhoexp}, we saw that as the amplitude of calibration
error fluctuations gets large the ratio between the rms of reconstructed map
residuals and the rms of the true ISW map, $s$, approaches a constant
value. Here we outline why this occurs.

Recall from Eq.~(\ref{eq:s_exp}) that our theoretical estimator $\langle s \rangle$ is written
\begin{equation}\label{eq:s_exp_reminder}
  \langle s\rangle = \frac{\sqrt{\langle \sigma_{\rm rec} \rangle^2 +\langle \sigma_{\rm ISW} \rangle^2   -  2\,\sum_{\ell i}\,(2\ell+1)\, R_{\ell}^i\tilde{C}_{\ell}^{{\rm ISW}-i}}}{ \langle \sigma_{\rm ISW} \rangle},
  \end{equation}
where
  \begin{align}
  \langle \sigma_{\rm ISW} \rangle &= \sqrt{ \sum_{\ell}\,(2\ell+1)\,\tilde{C}_{\ell}^{\rm ISW}}\text{, and}\\
  \langle \sigma_{\rm rec} \rangle &=\sqrt{ \sum_{\ell i j}\,(2\ell+1)\,R_{\ell}^iR_{\ell}^j\tilde{C}_{\ell}^{ij}}.\nonumber
\end{align}
In the case with a single LSS map, which we focus on here for simplicity, the reconstruction filter is
\begin{equation}
R_{\ell}^i = \frac{\Cl^{\rm gal-ISW}}{\Cl^{\rm gal}}.
\end{equation}
For clarity, and in contrast with the notation in the main text, here we
use tildes (as in $\tilde{C}_{\ell}$) to denote the $\Cl^{\rm true}$ which are associated with
observed or simulated maps. The $\Cl$ with no tilde will be the $\Cl^{\rm model}$
used to construct the ISW estimator.

Let us examine how the various terms scale as we increase the amplitude of
calibration errors. As the level of calibration errors---or any form of
noise---gets large,
\begin{equation}
\tilde{C}_{\rm \ell}^{\rm gal} \myeqtwo \Cl^{\rm noise} \propto A
\end{equation}
where $\Cl^{\rm noise}$ is the noise power spectrum and $A$
is a measure of its amplitude. The observed ISW power
$\tilde{C}_{\ell}^{\rm ISW}$ and
ISW-galaxy cross-power $\tilde{C}_{\ell}^{\rm gal-ISW}$ will not  depend on $A$.

For the calibration error studies in Sec.~\ref{sec:caliberror},
we focused on the case of residual calibration errors,
which are not accounted for in the ISW estimator.
In this scenario, any excess in observed power will be interpreted as a
bias and fit for via 
\begin{equation}
  \bar{b}^2\Cl^{\rm gal}=\tilde{C}_{\ell}^{\rm gal},
\end{equation}
according to the procedure described in 
Sec.~\ref{sec:biasfitting}.
Because $\Cl^{\rm gal}$ is independent of $A$, the resulting best fit value will be $\bar{b}^{\rm fit}\propto \sqrt{A}$.
The model   $\Cl(\bar{b}^{\rm fit})$  scales accordingly,
\begin{align}
  \Cl^{\rm gal}(\bar{b}^{\rm fit})&\propto A,\\
  \Cl^{\rm gal-ISW}(\bar{b}^{\rm fit})&\propto \sqrt{A},\\
  R_{\ell}&\propto \frac{1}{\sqrt{A}}.
\end{align}

Examining the terms in Eq.~(\ref{eq:s_exp_reminder}), we see that
$\langle \sigma_{\rm rec} \rangle$ and $\langle \sigma_{\rm ISW} \rangle$
will approach constants as $A$ grows, while the cross-term will go to zero
like $A^{-1/2}$. Thus,
in the case of unmodeled noise contributions to the galaxy maps, in the limit
of large noise,
\begin{equation}
  \langle s \rangle \myeqtwo \frac{\sqrt{\langle \sigma_{\rm rec} \rangle^2 +\langle \sigma_{\rm ISW} \rangle^2 }}{\langle \sigma_{\rm ISW} \rangle}.
\end{equation}
This is a constant greater than 1, in agreement with our results in
  the right panel of Fig.~\ref{fig:caltest_rhoexp}.

In contrast, if the $\Cl$ used in the ISW estimator correctly model the level of
galaxy noise---as occurs in the shot noise tests in
Sec.~\ref{sec:shottest}---the best fit bias parameter $\bar{b}^{\rm fit}$ will
remain close to 1.
In that case, the fact that noise is properly accounted for means that
\begin{equation}
  \Cl^{\rm gal}=\tilde{C}_{\rm ell}^{\rm gal}\propto A
\end{equation}
 while all other $\Cl$ and $\tilde{C}_{\ell}$ are independent of $A$. In this
case, as the noise power dominates over that of galaxies, the estimator
operator goes to zero according to 
\begin{equation}
  R_{\ell}\propto \frac{1}{A}.
\end{equation}
This means that for large levels of properly modeled noise, the
reconstructed map amplitude goes to zero. This causes $\langle \sigma_{\rm rec} \rangle $ and the cross-term in $\langle s\rangle$ to go to zero and so the
reconstruction  residuals are just a measure of the true ISW map:
\begin{equation}
  \langle s \rangle \myeqtwo \frac{\sqrt{\langle \sigma_{\rm ISW} \rangle^2 }}{\langle \sigma_{\rm ISW} \rangle} = 1.
\end{equation}

\bibliography{iswrec_paper2016_muirhuterer}{}
\end{document}